\documentclass[MSNbibl,nameyear,dvips]{arxstspdf}
\usepackage{flushend}
\usepackage{stfloats}
\usepackage{graphicx}
\volume{28}
\issue{4}
\pubyear{2013}
\firstpage{542}
\lastpage{563}
\doi{10.1214/13-STS441} 

\makeatletter

\newcommand{\rright}{\right}
\newcommand{\lleft}{\left}
\def\cal{\mathcal}

\newcommand{\pdiff}[2]{\frac{\partial #1}{\partial #2}}
\newcommand{\N}{\mathrm{N}}
\newcommand{\E}{\mathrm{E}}
\renewcommand{\P}{\mathbb{P}}
\def\IR{\mathbb{R}}
\makeatother

\begin{document}
\begin{frontmatter}

\title{Spatial and Spatio-Temporal Log-Gaussian Cox Processes:
Extending the Geostatistical Paradigm}
\runtitle{Spatial and Spatio-Temporal Log-Gaussian Cox Processes}

\begin{aug}
\author[a]{\fnms{Peter J.} \snm{Diggle}\corref{}\ead[label=e1]{p.diggle@lancaster.ac.uk}},
\author[b]{\fnms{Paula} \snm{Moraga}\ead[label=e2]{moragase@exchange.lancs.ac.uk}},
\author[c]{\fnms{Barry} \snm{Rowlingson}\ead[label=e3]{b.rowlingson@lancaster.ac.uk}}
\and
\author[d]{\fnms{Benjamin M.} \snm{Taylor}\ead[label=e4]{b.taylor1@lancaster.ac.uk}}
\runauthor{Diggle, Moraga, Rowlingson and Taylor}

\affiliation{Lancaster University Medical School and University of Liverpool}

\address[a]{Peter~J. Diggle is Distinguished University Professor, Lancaster University Medical
School, Lancaster, LA1 4YG, United Kingdom
and Professor, Institute of Infection and Global Health, University of
Liverpool, Liverpool L69 7BE, United Kingdom
\printead{e1}.}
\address[b]{\mbox{Paula~Moraga} is Research Associate, Lancaster University Medical
School, Lancaster, LA1 4YG, United Kingdom \printead{e2}.}
\address[c]{Barry~Rowlingson is Research Fellow, Lancaster University Medical
School, Lancaster, LA1 4YG, United Kingdom \printead{e3}.}
\address[d]{Benjamin~M. Taylor is Lecturer, Lancaster University Medical
School, Lancaster, LA1 4YG, United Kingdom \printead{e4}.}
\end{aug}

%
\begin{abstract}
In this paper we first describe the class of log-Gaussian Cox processes
(LGCPs) as models
for spatial and spatio-temporal point process data. We discuss
inference, with
a particular focus on the computational challenges of likelihood-based
inference.
We then demonstrate the usefulness of the LGCP by describing four
applications: estimating
the intensity surface of a spatial point process; investigating spatial
segregation
in a multi-type process; constructing spatially continuous maps of
disease risk from spatially
discrete data; and real-time health surveillance. We argue that
problems of this kind
fit naturally into the realm of geostatistics, which traditionally is
defined as
the study of spatially continuous processes using spatially discrete
observations
at a finite number of locations. We suggest that a more useful
definition of geostatistics
is by the class of scientific problems that it addresses, rather than by
particular models or
data formats.
\end{abstract}

%

\end{frontmatter}
%
\section{Introduction}\label{sec1}

Spatial statistics has been one of the most fertile areas for the
development of
statistical methodology during the second half of the twentieth
century. A striking, if
slightly contrived, illustration
of the pace of this development is the contrast between the 90 pages of
\citet{Bar75} and
the 900 pages of \citet{Cre91}. Cressie's book established a widely
used classification
of spatial statistics into three subareas: geostatistical data, lattice data,
spatial patterns (meaning point patterns). Within this classification,
geostatistical
data consist of observed values of some phenomenon of interest
associated with a set of spatial locations $x_i\dvtx i=1,\ldots,n$, where, in
principle, each $x_i$ could
have been any location $x$ within a
designated spatial region $A \subset\IR^2$. Lattice data consist of
observed values associated with a fixed set of locations $x_i\dvtx i=1,\ldots,n$, that is, the phenomenon
of interest exists only at those $n$ specific locations. Finally, in a
spatial pattern the data
are
a set of spatial locations $x_i\dvtx i=1,\ldots,n$ presumed to have been
generated as a partial
realisation of a point process that is itself the object of scientific interest.
Almost 20 years later, \citet{autokey31} used the same
classification but with a different terminology focused more on the underlying
process than on the extant data: continuous spatial variation,
discrete spatial variation, and spatial point processes. With this process-based
terminology in place, continuous spatial variation implies a stochastic
process $\{Y(x)\dvtx x \in\IR^2\}$, discrete spatial variation implies
only a finite-dimensional
random variable, $Y = \{Y_i\dvtx i=1,\ldots,n\}$, and a point pattern implies
a counting
measure, $\{dN(x)\dvtx x \in\IR^2\}$.

In this paper, we argue first
that the most important theoretical distinction within spatial
statistics is between
spatially continuous and spatially discrete stochastic processes, and
second that
most natural
processes are spatially continuous and should be modelled accordingly.
One consequence
of this point of view is that in many applications, maintaining a
one-to-one linkage
between data formats (geostatistical, lattice, point pattern) and
associated model classes
(spatially continuous, spatially discrete, point process) is
inappropriate. In particular,
we suggest a redefinition of geostatistics
as the collection of statistical models and methods whose purpose is to
enable predictive
inference about a spatially continuous, incompletely
observed phenome\-non, $S(x)$, say.

Classically,
geostatistical data
$Y_i\dvtx i=1,\ldots,n$ correspond to noisy versions of
$S(x_i)$. A standard geostatistical model, expressed here in hierarchical
form, is that ${\cal S} = \{S(x) \dvtx x \in\IR^2\}$ is a Gaussian
stochastic process, whilst conditional
on ${\cal S}$, the $Y_i$ are mutually independent, Normally distributed with
means $S(x_i)$ and common variance $\tau^2$. A second scenario, and the
focus of the current paper, is when ${\cal S}$ determines the
intensity, $\lambda(x)$, say,
of an
observed Poisson point process. An example that we will consider in detail
is a log-linear specification, $\lambda(x) = \exp\{S(x)\}$, where
${\cal S}$ is a Gaussian
process. A third form is when the
point process is reduced to observations of the numbers of points
$Y_i$ in each of $n$ regions $A_i$ that form a partition (or subset)
of the region of interest $A$. Hence, conditional on ${\cal S}$, the
$Y_i$ are mutually
independent, Poisson-distributed with means
%
%
\begin{equation}
\mu_i = \int_{A_i} \lambda(x) \,dx.
\label{eq:aggregated}
\end{equation}

In the remainder of the paper we show how the log-Gaussian Cox process
can be used in
a range of applications where $S(x)$ is
incompletely observed
through the lens of point pattern or aggregated count data. Sections~\ref{sec2}
to \ref{sect:computation} concern theoretical properties, inference and computation. Section~\ref{sec:applications} describes
several applications. Section~\ref{sec6} discusses the extension to
spatio-temporal data. Section~\ref{sec:synthesis}
gives an outline of how this approach to modelling incompletely
observed spatial phenomena
extends naturally to the joint analysis of multivariate spatial data
when the different
data elements are observed at incommensurate spatial scales. Section~\ref{sec8}
is a short, concluding discussion.

\section{The Log-Gaussian Cox Process}\label{sec2}

A (univariate, spatial)
Cox process (\cite{Cox55}) is a point process defined by the following two
postulates:
\begin{longlist}[CP2:]
\item[CP1:] $\Lambda= \{\Lambda(x)\dvtx x \in\IR^2\}$ is a nonnegative-valued
stochastic process;

\item[CP2:] conditional on the realisation $\Lambda(x) = \lambda(x)\dvtx\break   x \in\IR
^2$, the point
process is an inhomogeneous Poisson process with intensity $\lambda(x)$.
\end{longlist}

Cox processes are natural models for point process phenomena that are
environmentally driven,
much less natural for phenomena driven primarily by interactions
amongst the points. Examples
of these two situations in an epidemiological context would be the
spatial distribution of
cases of a noninfectious or infectious disease, respectively. In a
noninfectious disease, the
observed spatial pattern of cases results from spatial variation in the
exposure of
susceptible individuals to a combination of observed and unobserved
risk-factors. Conditional
on exposure, cases occur independently. In contrast,
in an infectious disease the observed pattern is at least partially the
result of infectious
cases transmitting the disease to nearby susceptibles. Notwithstanding this
phenomenological
distinction, it can be difficult, or even impossible, to distinguish
empirically between
processes representing stochastically
independent variation in a heterogeneous environment
and stochastic interactions in a homogeneous environment (\cite{Bar64}).

The moment properties of a Cox process are inherited from those of the
process $\Lambda(x)$.
For example, in the stationary case the intensity of the Cox process is equal
to the expectation of $\Lambda(x)$ and the covariance density of the
Cox process is equal
to the covariance function of $\Lambda(x)$. Hence, writing $\lambda
=\mathrm{E}[\Lambda(x)]$
and $C(u) =  \operatorname{Cov}\{\Lambda(x),\Lambda(x-u)\}$, the reduced second moment
measure or $K$-function (Ripley \citeyear{Rip76}, \citeyear{Rip77}) of the Cox process is
%
%
\begin{equation}
K(u) = \pi u^2 + 2 \pi\lambda^{-2} \int
_0^u C(v) v \,dv. \label{eq:cox_K}
\end{equation}

M{\o}ller, Syversveen and Waagepetersen (\citeyear{MllSyvWaa98}) introduced the class of
log-Gaussian
processes\break  (LGCPs). As the name implies, an LGCP is a Cox process with
$\Lambda(x) = \exp\{S(x)\}$, where ${\cal S}$ is a Gaussian process.
This construction
has an elegant simplicity. One of its attractive features is that the
trac\-tability of
the multivariate Normal distribution carries over, to some extent, to
the associated
Cox process.

In the stationary
case, let $\mu= \mathrm{E}[S(x)]$ and $C(u) = \sigma^2 r(u) = \operatorname{Cov}\{
S(x),S(x-u)\}$.
It follows from the moment properties of the log-Normal distribution
that the associated
LGCP has intensity $\lambda=\break \exp(\mu+ 0.5 \sigma^2)$ and covariance density
$g(u) =\break  \lambda^2[\exp\{\sigma^2 r(u)\}-1]$. This makes it both
convenient and natural
to re-parameterise the model as
%
%
\begin{equation}
\Lambda(x) = \exp\bigl\{\beta+S(x)\bigr\}, \label{eq:param}
\end{equation}
where $\mathrm{E}[S(x)] = - 0.5 \sigma^2$, so that $\mathrm{E}[\exp\{S(x)\}]=1$
and $\lambda= \exp(\beta)$.
This
re-parameterisation gives a clean separation between first-order (mean
value) and
second-order (variation about the mean) properties. Hence, for example,
if we wished to
model a spatially varying intensity by including one or more spatially
indexed explanatory
variables $z(x)$, a natural first approach would be to retain the
stationarity of $S(x)$ but
replace the constant intensity $\lambda$ by a regression model,
$\lambda(x) = \lambda\{z(x); \beta\}$. The resulting Cox process is now
an intensity-reweighted
stationary point process (Baddeley, M{\o}ller and Waagepeter\-sen, \citeyear{BadMllWaa00}),
which is the
analogue of a real-valued process with a spatially varying mean and a
stationary residual.

The definition of a multivariate LGCP is imme\-diate---we simply replace
the scalar-valued Gaussian process $S(x)$ by a vector-valued
multivariate
Gaussian process---and its moment properties are equally tractable.
For example, if $S(x)$
is a stationary bivariate Gaussian process with intensities
$\lambda_1$ and $\lambda_2$, and cross-covariance function
$C_{12}(u) = \sigma_1 \sigma_2 r_{12}(u)$, the cross-covariance density
of the
associated Cox process is
$g_{12}(u) = \lambda_1 \lambda_2 [\exp\{\sigma_1 \sigma_2 r_{12}(u)\}-1]$.

There is an extensive literature on parametric specifications for the covariance
structure of real-valued processes $S(x)$; for a recent summary, see
\citet{GneGut10N1}. The theoretical requirement for a function
$C(x,y)$ to be a valid covariance function is that it be
positive-definite, meaning
that for all positive integers $n$, any associated set of points
$x_i \in\IR^2\dvtx i=1,\ldots,n$, and any associated set of real numbers
$a_i\dvtx i=1,\ldots,n$,
%
%
\begin{equation}
\sum_{i=1}^n \sum
_{j=1}^n a_i a_j
C(x_i,x_j) \geq0. \label{eq:posdef}
\end{equation}
Checking that (\ref{eq:posdef}) holds for an arbitrary
candidate $C(x,y)$ is not straightforward.
In practice, we choose covariance functions from a catalogue of
parametric families that
are known to be valid. In the stationary case, a widely used family is the
\citet{M60} class $C(u) = \sigma^2 r(u; \phi, \kappa)$, where
%
%
\begin{eqnarray}\label{eq:matern}
&&r(u; \phi, \kappa)
\nonumber
\\[-8pt]
\\[-8pt]
\nonumber
&&\quad = \bigl\{2^{\kappa-1} \Gamma(\kappa)\bigr
\}^{-1} (u/\phi )^\kappa K_\kappa(u/\phi)\quad u \geq0.
\end{eqnarray}
In (\ref{eq:matern}), $\Gamma(\cdot)$ is the complete Gamma function,
$K_\kappa(\cdot)$ is
a modified Bessel function of order $\kappa$, and $\phi>0$ and $\kappa
>0$ are parameters. The
parameter $\phi$ has units of distance,
whilst $\kappa$ is a dimensionless shape parameter
that determines the differentiability of the corresponding Gaussian
process; specifically,
the process is $k$-times
mean square differentiable if $\kappa> k$. This physical
interpretation of $\kappa$ is useful because $\kappa$
is difficult to estimate empirically (\cite{Zha04}), hence,
a widely used strategy is to choose between a small set of values
corresponding to different
degrees of differentiability, for example, $\kappa= 0.5, 1.5$ or $2.5$.
Estimation of $\phi$ is more straightforward.

In summary, the LGCP is the natural analogue for point process data of
the linear Gaussian
model for real-valued geostatistical data (\cite{DigRib07}).
Like the linear Gaussian
model, it lacks any mechanistic interpretation. Its principal virtue is that
it provides a flexible and
relatively tractable class of empirical models for describing spatially
correlated phenomena.
This makes it extremely useful in a range of applications where the
scientific focus is
on spatial prediction rather than on testing mechanistic hypotheses.
Section~\ref{sec:applications} gives several
examples.

\section{Inference for Log-Gaussian Cox Processes}\label{sec3}

In this section we distinguish between two inferential targets, namely,
estimation
of model parameters and prediction of the realisations of unobserved stochastic
processes. Within the Bayesian paradigm, this distinction is often
blurred, because
parameters are treated as unobserved random variables and the
formal machinery of inference is the same in both cases, consisting of
the calculation of
the conditional distribution of the target given the data. However,
from a
scientific perspective parameter estimation and prediction are
fundamentally different,
because the former concerns properties of the process being modelled whereas
the latter concerns
properties of a particular \textit{realisation} of that process.

\subsection{Parameter Estimation}\label{sec3.1}

For parameter estimation, we consider three approaches: moment-based
estimation, maximum likelihood estimation, and Bayesian estimation. The
first approach is
typically very simple to implement and is useful for the initial
exploration of candidate models.
The second and third are more principled, both being likelihood-based.

\subsubsection{Moment-based estimation}\label{sec3.1.1}

In the stationary case, moment-based estimation consists of minimising
a measure of
the discrepancy between empirical and theoretical second-moment
properties. One class
of such measures is a weighted least squares criterion,
%
%
\begin{equation}
D(\theta) = \int_0^{u_0} w(u) \bigl\{
\hat{K}(u)^c - K(u;\theta)^c\bigr\}^2 \,du.
\label{eqn:moment-based}
\end{equation}
In the intensity-re-weighted case, (\ref{eqn:moment-based}) can still
be used after
separately estimating a regression model for a spatially varying
$\lambda(x)$ under the
working assumption that the data are a partial realisation of an
inhomogeneous Poisson
process.

This method of estimation has an obviously
\textit{ad hoc} quality. In particular, it is difficult to give generally
applicable guidance
on appropriate choices for the values of $u_0$ and $c$ in (\ref
{eqn:moment-based}). Because
the method is intended only to give preliminary estimates, there is
something to be said
for simply matching $\hat{K}(u)$ and $K(u;\theta)$ by eye. The \texttt{R}
(\cite{Tea13}) package
\texttt{lgcp} (Taylor {et al.}, \citeyear{Tayetal}) includes an interactive graphics
function to facilitate
this.

\subsubsection{Maximum likelihood estimation}\label{sec3.1.2}

The general form of the Cox process likelihood associated with
data $X = \{x_i \in A\dvtx i=1,\ldots,n\}$ is
%
%
\begin{eqnarray}\label{cox1}
\ell(\theta;X)&=&\mathrm{P}(X|\theta)=\int_\Lambda \mathrm{P}(X,
\Lambda|\theta)\,d\Lambda
\nonumber
\\[-8pt]
\\[-8pt]
\nonumber
&=&\mathrm{E}_{\Lambda|\theta}\bigl(\ell ^*(\Lambda;X)\bigr),
\end{eqnarray}
where
%
%
\begin{equation}
\ell^*(\Lambda;X) = \prod_{i=1}^n
\Lambda(x_i) \biggl\{\int_A \Lambda(x) \,dx
\biggr\}^{-n} \label{cox0}
\end{equation}
is the likelihood for an inhomogeneous Poisson process with intensity
$\Lambda(x)$.
The evaluation of (\ref{cox1}) involves integration over the
infinite-dimensional distribution of $\Lambda$.
In Section~\ref{sec4.1} below we describe an implementation in which the continuous
region of interest $A$ is approximated by a finely spaced regular
lattice, hence replacing
$\Lambda$ by a finite set of values $\Lambda(g_k)\dvtx k=1,\ldots,N$, where the points
$g_1,\ldots,g_N$ cover $A$. Even so, the high dimensionality of the
implied integration appears to present a formidable obstacle to
analytic progress.
One solution,
easily stated but hard to implement robustly and efficiently, is to use
Monte Carlo methods.

Monte Carlo evaluation of (\ref{cox1}) consists of approximating the
expectation by an
empirical average over simulated realisations of some kind. A crude
Monte Carlo method would use
the approximation
%
%
\begin{equation}
\ell_{\mathrm{MC}}(\theta) = s^{-1} \sum
_{j=1}^s \ell\bigl(\theta;X,\lambda^{(j)}
\bigr), \label{cox2}
\end{equation}
where $\lambda^{(j)} = \{\lambda^{(j)}(g_k)\dvtx k=1,\ldots,N\}\dvtx j=1,\ldots,s$ are
simulated realisations of $\Lambda$ on
the set of grid-points $g_k$. In practice, this is hopelessly inefficient.
A better approach is to use an ingenious method due to \citet{Gey99}, as follows.

Let $f(X,\Lambda;\theta)$ denote the un-normalised joint density of $X$ and
$\Lambda$.
Then, the associated likelihood is
%
%
\begin{equation}
\ell(\theta;X,\Lambda) = f(X,\Lambda;\theta)/a(\theta), \label{eq:cox1}
\end{equation}
where
%
%
\begin{equation}
a(\theta) = \int f(X,\Lambda;\theta) \,d \Lambda \,dX \label{eq:cox2}
\end{equation}
is the intractable normalising constant for $f(\cdot)$.
It follows that
%
%
\begin{eqnarray}
\label{eqn:cox3} &&\mathrm{E}_{\theta_0}\bigl[f(X,\Lambda;\theta)/f(X,\Lambda;
\theta_0)\bigr] \nonumber\\
&&\quad =  \int\!\!\int f(X,\Lambda;\theta)/f(X,\Lambda;
\theta_0) \nonumber\\
&&\hspace*{42pt}{}\times\frac
{f(X,\Lambda;\theta_0)}{a(\theta_0)} \,d \Lambda \,dX
\\
&&\quad =  \frac{1}{a(\theta_0)} \int f(X,\Lambda;\theta) \,d \Lambda \,dX
\nonumber\\
&&\quad =  a(\theta)/a(\theta_0),\nonumber
\end{eqnarray}
where $\theta_0$ is any convenient, fixed value of $\theta$, and $\mathrm{E}_{\theta_0}$ denotes
expectation when $\theta=\theta_0$.
However, the function
$f(X,\Lambda;\theta)$ in (\ref{eq:cox1}) is also an un-normalised conditional
density for $\Lambda$ given $X$. Under this
second interpretation, the corresponding normalised
conditional density is $f(X,\Lambda;\theta)/a(\theta|X) $, where
%
%
\begin{equation}
a(\theta|X) = \int f(X,\Lambda;\theta) \,d\Lambda, \label{eq:aratio2}
\end{equation}
and the same argument as before gives
%
%
\begin{eqnarray}\label{eqn:cox4}
&&\mathrm{E}_{\theta_0}\bigl[f(X,\Lambda;\theta)/f(X,\Lambda;
\theta_0)|X\bigr]
\nonumber
\\[-8pt]
\\[-8pt]
\nonumber
&&\quad= a(\theta|X)/a(\theta_0|X).
\end{eqnarray}
It follows from (\ref{cox1}), (\ref{eq:cox1}) and (\ref{eq:aratio2})
that the likelihood for the observed data, $X$, can be written as
%
%
\begin{equation}\quad
\ell(\theta;X) = \int\frac{f(x,\Lambda;\theta)}{a(\theta)} \,d \Lambda= a(\theta|X)/a(\theta).
\label{eqn:ell1}
\end{equation}
Hence, the log-likelihood ratio between any two parameter values,
$\theta$ and $\theta_0$, is
%
%
\begin{eqnarray}
\label{eqn:ell2}&& L(\theta;X) - L(\theta_0;X) \nonumber\\
&&\quad =  \log\bigl\{a(
\theta|X)/a(\theta)\bigr\} - \log\bigl\{a(\theta_0|X)/a(
\theta_0)\bigr\}
\\
&&\quad =  \log\bigl\{a(\theta|X)/a(\theta_0|X)\bigr\} - \log\bigl\{a(
\theta)/a(\theta_0)\bigr\}.\nonumber
\end{eqnarray}
Substitution from (\ref{eqn:cox3}) and (\ref{eqn:cox4}) gives the
result that
%
%
\begin{eqnarray} \label{eqn:coxLR}
&& L(\theta;X) - L(\theta_0;X)
\nonumber
\\
&&\quad= \log\mathrm{E}_{\theta_0}\bigl[r(X,
\Lambda,\theta ,\theta_0)|X\bigr]\\
&&\qquad{} - \log\mathrm{E}_{\theta_0}
\bigl[r(X,\Lambda,\theta,\theta_0)\bigr],\nonumber
\end{eqnarray}
where $r(X,\Lambda,\theta,\theta_0) = f(X,\Lambda;\theta)/f(X,\Lambda
;\theta_0)$.
For any fixed value of $\theta_0$, a
Monte Carlo approximation to the log-likelihood, ignoring the constant
term $L(\theta_0)$
on the left-hand side of (\ref{eqn:coxLR}), is therefore given by
%
%
\begin{eqnarray}\label{eqn:coxMCLR}
\hat{L}(\theta) &= &\log \Biggl\{s^{-1}\sum
_{j=1}^s r\bigl(X,\lambda^{(j)},\theta,
\theta_0\bigr) \Biggr\}
\nonumber
\\[-8pt]
\\[-8pt]
\nonumber
&&{}- \log \Biggl\{s^{-1}\sum
_{j=1}^s r\bigl(X^{(j)},
\lambda^{(j)},\theta,\theta _0\bigr) \Biggr\}.
\end{eqnarray}
The result (\ref{eqn:coxMCLR}) provides a Monte Carlo approximation
to the log-likelihood function, and therefore to the maximum likelihood
estimate $\hat{\theta}$,
by simulating the process only at a single value, $\theta_0$.
The
accuracy of the approximation depends on the number of simulations,
$s$, and on
how close $\theta_0$ is to $\hat{\theta}$.

Note that in the second term on the right-hand side of (\ref
{eqn:coxMCLR}) the pairs
$(X^{(j)}, \lambda^{(j)})$ are simulated joint realisations of $X$ and
$\Lambda$
at $\theta=\theta_0$, whilst in the first term $X$ is held fixed at the
observed data
and the simulated realisations $\lambda^{(j)}$ are conditional on $X$.
Conditional simulation
of $\Lambda$ requires Markov chain Monte Carlo (MCMC)
methods, for which careful tuning is generally
needed. We discuss computational issues, including
the design of a suitable MCMC algorithm, in Section~\ref{sect:computation}.

\subsubsection{Bayesian estimation}\label{sec3.1.3}

One way to implement Bayesian estimation would be directly to combine
Monte Carlo evaluation
of the likelihood with a prior for $\theta$. However, it turns out to
be more efficient to
incorporate Bayesian estimation and prediction into a single MCMC
algorithm, as described
in Section~\ref{sect:computation}.

\subsection{Prediction}\label{sec3.2}

For prediction, we consider
plug-in and Bayesian prediction. Suppose, quite generally, that data
$Y$ are to be used to predict
a target $T$ under an assumed model with parameters $\theta$. Then,
plug-in prediction consists
of a series of probability statements within the conditional
distribution $[T|Y;\hat{\theta}]$,
where $\hat{\theta}$ is a point estimate of $\theta$, whereas Bayesian
prediction replaces
$[T|Y;\hat{\theta}]$ by
%
%
\begin{equation}
[T|Y] = \int[T|Y;\theta] [\theta|Y] \,d \theta. \label{eq:bayes_prediction}
\end{equation}
This
shows that
Bayesian prediction is a weighted average of plug-in predictions, with different
values of $\theta$ weighted according to
the Bayesian posterior for $\theta$. The Bayesian solution (\ref
{eq:bayes_prediction}) is the
more correct in that it incorporates parameter uncertainty in a way
that is both natural,
albeit on
its own terms, and elegant.

\section{Computation}\label{sect:computation}

Inference for LGCPs is a computationally challenging problem.
Throughout this section we
will use the notation and language of purely spatial processes on $\IR
^2$, but the discussion applies in more general settings including
spatio-temporal LGCPs.

\subsection{The Computational Grid}\label{sec4.1}

Although we model the latent process
$S$ as a spatially continuous process, in practice, we work with
a piecewise-constant equivalent to the LGCP model
on a collection of cells that form a disjoint partition of the
region of interest, $A$.
In the limit as the number of cells tends to infinity, this
process behaves like its spatially
continuous counterpart. We call the collection of cells on which we
represent the process the \textit{computational grid}. The choice of grid
reflects a balance between
computational complexity and accuracy of approximation.
The computational bottleneck
arises through the need to invert the covariance matrix, $\Sigma$, corresponding
to the variance of $S$ evaluated on the computational grid.

Typically, we shall use a computational grid of square cells. This is
an example of a
\textit{regular grid},
by which we mean that on an extension of the grid notionally wrapped on
a torus, a strictly stationary covariance function of the process
on $\IR^2$ will induce a block-circulant covariance structure on the
grid (\cite*{WooCha94}; M{\o}ller, Syversveen and Waagepetersen,
\citeyear{MllSyvWaa98}). For simplicity of presentation, we make no distinction between
the extended grid and the original grid, since for extensions that at
least double the width and height of the original grid, the toroidal
distance between any two cells in the original observation window
coincides with their Euclidean distance in $\IR^2$. For a second-order
stationary process $S$, inversion of $\Sigma$ on a regular grid is best
achieved using Fourier methods (\cite{FriJoh}). On irregular
grids, sparse matrix methods in conjunction with an assumption of
low-order Markov dependence are more efficient
(\cite{RueHel05}; \cite{RueMarCho09}; Lindgren, Rue and Lindstr\"{o}m, \citeyear{LinRueLin11}).
In this context,
Lindgren, Rue and Lindstr\"{o}m (\citeyear{LinRueLin11})
demonstrate a link between models assuming a Markov
dependence structure and spatially continuous models whose
covariance function belongs to a restricted subset of the Mat\'ern class.

\subsection{Implementing Bayesian Inference, MCMC or INLA?}\label{sec4.2}

We now suppose that the computational grid has been defined and
the point process data $X$ have been converted to a set of counts, $Y$, on
the grid cells; note that we envisage using a finely spaced grid, for
which cell-counts will
typically be 0 or 1. Our goal is to use the data $Y$ to make inferences about
the latent process $S$ and the parameters $\beta$ and $\theta$, which,
respectively,
parameterise the intensity of the LGCP and the covariance structure of
$S$.

In the Bayesian paradigm we treat $S$, $\beta$ and $\theta$ as random
variables, assign
priors to the model parameters $(\beta, \theta)$
and make inferential statements using the posterior/predictive distribution,
\[
[S,\beta,\theta|Y] \propto[Y|S,\beta,\theta] [S|\theta] [\beta,\theta].
\]
Two options for computation are as follows: MCMC, which
generates random samples from $[S,\beta,\theta|Y]$, and the integrated
nested Laplace approximation (INLA), which uses a mathematical approximation.

\citet{TayDigN1} compare the performance of MCMC and INLA
for a spatial LGCP with constant expectation $\beta$
and parameters $\theta$ treated as known values.
In this restricted scenario, they found that MCMC, run for 100,000 iterations,
delivered more accurate estimates of predictive probabilities
than INLA. However, they acknowledged
that ``further research is required in order to design better
MCMC algorithms that also provide inference for the parameters of the
latent field''.

Approximate methods such as INLA have the advantages that they produce
results quickly and circumvent the need to assess the convergence and
mixing properties
of an MCMC algorithm. This makes INLA very convenient for quick
comparisons amongst
multiple candidate models, which would be a daunting task for MCMC.
Against this,
MCMC methods are more flexible in that
extensions to standard classes of models can usually be accommodated
with only a modest amount of coding effort. Also, an important consideration
in some applications is that the currently available
software implementation of
INLA is limited to the evaluation of predictive distributions for
univariate, or, at best, low-dimensional, components
of the underlying model, whereas MCMC provides direct
access to joint posterior/predictive distributions
of nonlinear functions of the parameters and of the latent process $S$.
Mixing INLA and MCMC can therefore be a good overall computational
strategy. For example, \citet{HarTie} use a heavy-tailed approximation
similar in spirit to INLA to construct efficient MCMC proposal schemes.

\subsubsection{Markov Chain Monte Carlo inference for log-Gaussian Cox
processes}\label{sec4.2.1}

MCMC methods generate
samples from a Markov chain whose stationary distribution is the target
of interest, in our case $[S,\beta,\break \theta|Y]$.
Such samples are inherently dependent but, subject to careful checking of
mixing and convergence properties, their empirical distribution is an
unbiased estimate
of the target, and, in principle, the associated
Monte Carlo error
can be made arbitrarily small by using a sufficiently long run of the chain.
In the current context,
we follow M{\o}ller, Syversveen and Waagepetersen (\citeyear{MllSyvWaa98}) and \citet{BriDig01}
in using a standardised version of $S$, denoted $\Gamma$, and transform
$\theta$ to the log-scale, so that the MCMC algorithm operates
on the whole of $\IR^d$, rather than on a restricted subset. We
denote the $i$th sample from the chain by $\zeta^{(i)}$ and write
$\pi(\zeta|Y)$ for the target distribution.

The aim in designing MCMC algorithms for any specific class of problems
is to achieve faster
convergence and better mixing than would be obtained by generic
off-the-shelf methods.
Gilks, Richardson and Spiegelhalter (\citeyear{GilRicSpi95})
and \citet{GamLop06} give overviews of the extensive
literature on
this topic.
We focus our discussion on the Metropolis-Hastings (MH) algorithm, which
includes as a special case the popular Gibbs sampler
(Metropolis\break  {et~al.}, \citeyear{Metetal53}; \cite{Has70}; \cite{GemGem84};
Spiegelhalter, Thomas and Best, \citeyear{SpiThoBes}).
In order to use the MH algorithm, we require a proposal density, $q(
\cdot|\zeta^{(i-1)})$. At the $i$th iteration of the algorithm, we
sample a candidate, $\zeta^{(i^*)}$, from $q(\cdot)$, and set $\zeta
^{(i)}=\zeta^{(i^*)}$ with probability
\[
\min \biggl\{1,\frac{\pi(\zeta^{(i^*)}|Y)}{\pi(\zeta^{(i-1)}|Y)}\frac
{q(\zeta^{(i-1)}|\zeta^{(i^*)})}{q(\zeta^{(i^*)}|\zeta^{(i-1)})} \biggr\},
\]
otherwise set $\zeta^{(i)}=\zeta^{(i-1)}$.
The choice of $q(\cdot)$ is critical. Previous research
on inferential methods for spatial and spatio-temporal log-Gaussian Cox
processes
has advocated the Metropolis-adjusted Lan\-gevin algorithm (MALA), which
mimics a Langevin diffusion on the target of interest;
see \citet{RobTwe96},
M{\o}ller, Syversveen and Waagepeter\-sen (\citeyear{MllSyvWaa98}) and \citet{BriDig01}; note also
\citet{BriDig03} and \citet{TayDigN2}.
Alternatives to MH include Hamiltonian Monte Carlo methods,
as discussed in \citet{GirCal11}.

The Metropolis-adjusted Langevin algorithm exploits gradient
information to
identify efficient proposals. The algorithms in this article make use
of a ``pre-conditioning matrix'', $\Xi$ (\cite{GirCal11}),
to define
the proposal
%
%
\begin{eqnarray}
\quad&& q\bigl(\zeta^{(i^*)}|\zeta^{(i-1)}\bigr)
\nonumber
\\
&&\quad= \N \biggl[
\zeta^{(i^*)};\\
&&\hspace*{2pt}\quad\qquad{}\zeta ^{(i-1)}+\frac{h^2}2\Xi\nabla\log\bigl\{
\pi\bigl(\zeta^{(i-1)}|Y\bigr)\bigr\},h^2\Xi \biggr],\nonumber
\end{eqnarray}
where $h$ is a scaling constant. Ideally, $\Xi$ should
be the negative inverse of the Fisher information matrix evaluated at
the maximum likelihood estimate of $\zeta$, that is, $\Xi_{\mathrm{opt}} =
\{-\E[\mathcal I(\hat\zeta)]\}^{-1}$
where $\mathcal I$ is the observed information. However, this matrix
is massive, dense and intractable.
In practice, we can obtain an efficient algorithm by choosing $\Xi$ to
be an approximation of $\Xi_{\mathrm{opt}}$ and further by changing $h$
during the course of the algorithm
using adaptive MCMC (\cite{AndTho08}; \cite{RobRos07}). In MALA algorithms, $h$ can be tuned
adaptively to achieve an approximately optimal acceptance rate of 0.574
(\cite{RobRos01}).

Since the gradient of $\log\pi$ with respect to $\theta$ can be both
difficult to compute
and computationally costly, we instead suggest a random walk proposal for
the $\theta$-component of $\zeta$. In the examples described in
Section~\ref{sec:applications}
we used the following overall proposal:
%
%
\begin{eqnarray}\label{eqn:overall}
&&q\bigl(\zeta^{(i^*)}|\zeta^{(i-1)}\bigr)\nonumber\\
&&\quad=\N\lleft[
\zeta^{(i^*)};\phantom{\pmatrix{
\displaystyle\frac{h^2h_{\Gamma}^2}{2}\Xi_\Gamma\pdiff {\log\bigl\{\pi
\bigl(\zeta^{(i-1)}|Y\bigr)\bigr\}} {\Gamma}
\vspace*{2pt}\cr
+\displaystyle\frac{h^2h_{\beta}^2}{2}\Xi_\beta\pdiff{\log\bigl\{\pi
\bigl(\zeta ^{(i-1)}|Y\bigr)\bigr\}} {\beta}
\vspace*{2pt}\cr
\theta^{(i-1)} }}\right.
\nonumber
\\[-8pt]
\\[-8pt]
\nonumber
&&\hspace*{36pt}\left.\pmatrix{
\Gamma^{(i-1)}+\displaystyle\frac{h^2h_{\Gamma}^2}{2}\Xi_\Gamma\pdiff {\log\bigl\{\pi
\bigl(\zeta^{(i-1)}|Y\bigr)\bigr\}} {\Gamma}
\vspace*{2pt}\cr
\beta^{(i-1)}+\displaystyle\frac{h^2h_{\beta}^2}{2}\Xi_\beta\pdiff{\log\bigl\{\pi
\bigl(\zeta ^{(i-1)}|Y\bigr)\bigr\}} {\beta}
\vspace*{2pt}\cr
\theta^{(i-1)} },\right.\\
&&\hspace*{-67pt}\left.\phantom{\pmatrix{
\displaystyle\frac{h^2h_{\Gamma}^2}{2}\Xi_\Gamma\pdiff {\log\bigl\{\pi
\bigl(\zeta^{(i-1)}|Y\bigr)\bigr\}} {\Gamma}
\vspace*{2pt}\cr
+\displaystyle\frac{h^2h_{\beta}^2}{2}\Xi_\beta\pdiff{\log\bigl\{\pi
\bigl(\zeta ^{(i-1)}|Y\bigr)\bigr\}} {\beta}
\vspace*{2pt}\cr
\theta^{(i-1)} }} h^2
\pmatrix{ h_{\Gamma}^2
\Xi_\Gamma& 0 & 0
\vspace*{2pt}\cr
 0 & h_{\beta}^2\Xi_\beta& 0
\vspace*{2pt}\cr
0 & 0 & ch_{\theta}^2\Xi_\theta}
 \rright].\nonumber
\end{eqnarray}
In (\ref{eqn:overall}), $\Xi_\Gamma$ is an approximation to $\{-\E
[\mathcal I(\hat\Gamma)]\}^{-1}$, and similarly for $\Xi_\beta$ and $\Xi
_\theta$. The constants $h_{\Gamma}^2$, $h_{\beta}^2$ and $h_{\theta
}^2$ are the approximately optimal scalings for Gaussian targets
explored by the Gaussian random walk or MALA proposals (\cite{RobRos01});
these are, respectively, $1.65^2/\dim(\Gamma)^{1/3}$,\break
$1.65^2/\dim(\beta)^{1/3}$ and $2.38^2/\dim(\theta)$, where $\dim$ is
the dimension.

The acceptance rate for
a random walk proposal is often tuned to around 0.234,
which is optimal for a Gaussian target in the limit as the dimension of
the target goes to infinity.
At each step in our algorithm, we jointly propose new
values for $(S,\beta)$ and for $\theta$ using, respectively, a MALA
and a random walk
component in the overall
proposal, but we also seek to maintain an
acceptance rate of 0.574 to achieve optimality for the MALA parts of
the proposal. As a
compromise, in our proposal we
scale the matrix $\Xi_\theta$ by a constant factor $c$ and
the proposal covariance matrix by a single adaptive $h$.
In the examples described in Section~\ref{sec:applications}
we used a value of $c=0.4$, which appears to work well across a range
of scenarios.

\section{Applications} \label{sec:applications}

\subsection{Smoothing a Spatial Point Pattern} \label{sec:smoothing}

\begin{figure}[b]

\includegraphics{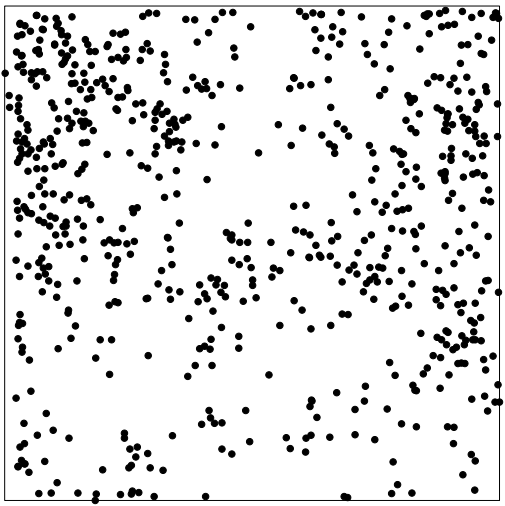}

\caption{Locations of 703 hickories in a 19.6 acre square plot,
re-scaled to 100 by 100 units (\cite{Ger69}).}\label{fig:trees}
\end{figure}

The intensity, $\lambda(x)$, of an inhomogeneous spatial point process
is the
unique nonnegative valued function such
that the expected number of points of the process, called
\textit{events}, that fall within any
spatial region $B$ is
%
%
\begin{equation}
\mu(B) = \int_B \lambda(x) \,dx. \label{eq:intensity}
\end{equation}
Suppose that we wish to estimate $\lambda(x)$ from a partial realisation
consisting of all of the events of the process that fall within
a region $A$, hence, $X = \{x_i \in A\dvtx i=1,\ldots,n\}$. Figure~\ref{fig:trees}
shows an example in which the data are the locations of 703 hickory
trees in a
19.6 acre (281.6 by 281.6 metre) square region $A$ (\cite{Ger69}), which
we have re-scaled to be of dimension 100 by 100.

\begin{figure*}

\includegraphics{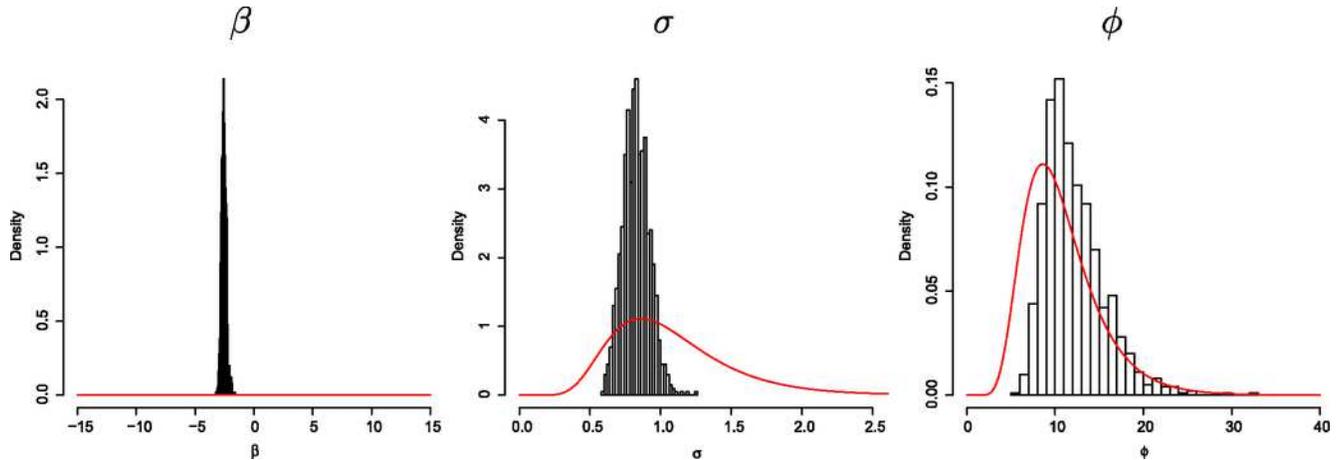}

\caption{Prior (continuous curve) and posterior (histogram)
distributions for the parameters $\beta$,
$\sigma$ and $\phi$ in the LGCP model for the hickory data.}
\label{fig:hickory_estimates}
\end{figure*}

An intuitively reasonable class
of estimators for $\lambda(x)$ is obtained by counting the number of
events that lie
within some fixed distance, $h$, say, of $x$ and dividing by $\pi h^2$
or, to allow for
edge-effects, by the area, $B(x,t)$, of the intersection of $A$ and a
circular disc with centre $x$ and radius $h$, hence,
%
%
\begin{equation}
\tilde{\lambda}(x;h) = B(x;h)^{-1} \sum_{i=1}^n
I\bigl(\Vert x-x_i\Vert \leq h\bigr). \label{eq:kernel}
\end{equation}
This estimate is, in essence, a simple form of bivariate kernel
smoothing with a
uniform kernel function
(\cite{Sil86}). \citet{BerDig89} derived the mean square
error of
(\ref{eq:kernel}) as a function of $h$ under the assumption that the
underlying point
process is a stationary
Cox process. They then showed how to estimate, and thereby
approximately minimise,
the mean square error without further parametric assumptions.

A different way to formalise the smoothing problem is as a prediction
problem associated
with the log-Gaussian Cox process, (\ref{eq:param}).
In this formulation, $\Lambda(x) = \exp\{\beta+ S(x)\}$, where
$S(\cdot)$ is a stationary Gaussian process indexed by a parameter
$\theta$ and the
target for prediction is $\Lambda(x)$. The formal solution is the
predictive distribution
of $\Lambda(\cdot)$ given $X$.
For a smooth estimate,
analogous to (\ref{eq:kernel}), we take $\hat{\lambda}(x)$ to be a suitable
summary of the predictive distribution, for example, its
point-wise expectation or median.
This is still a nonparametric solution, in the sense that no parametric
form is specified
in advance
for $\hat{\lambda}(x)$. The parameterisation of the Gaussian process
$S(\cdot)$ is the counterpart
of the choices made in the kernel estimation approach, namely, the
specification of the
uniform kernel in (\ref{eq:kernel})
and the value of the bandwidth, $h$.

\begin{figure*}[b]

\includegraphics{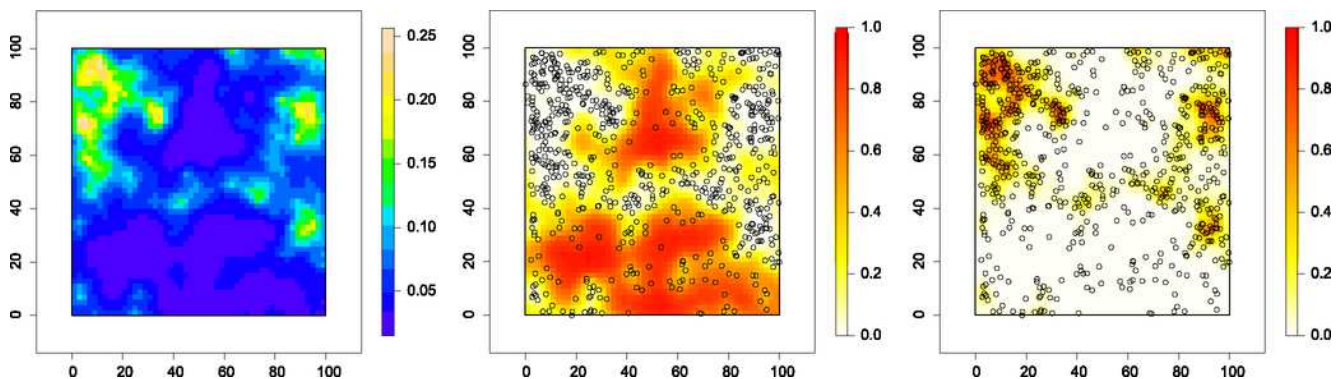}

\caption{Left: 50\% posterior percentiles of $\Lambda(x) = \exp\{\beta
+S(x)\}$ for the hickory data.
Middle: plot of posterior $\P\{\exp[S(x)]<1/2\}$. Right: plot of
posterior $\P\{\exp[S(x)]>2\}$. Middle and right plots also show the
locations of the trees.}
\label{fig:hickory_surfaces}
\end{figure*}

For this application, we specify that $S(\cdot)$ has mean $- 0.5 \sigma^2$,
variance $\sigma^2$ and
exponential correlation function, $r(u)=\exp(-u/\phi)$, hence, $\theta
= (\sigma^2,\phi)$.
We conduct Bayesian
predictive inference using MCMC\break  methods
implemented in an extension of the \texttt{R} package \texttt{lgcp}
(Taylor {et al.}, \citeyear{Tayetal}). For
$\beta$ we chose a diffuse prior, $\beta\sim N(0, 10^6)$.
For $\sigma$ and $\phi$, we chose Normal priors on the log scale: $\log
\sigma\sim N(\log(1), 0.15)$ and\ $\log\phi\sim N(\log(10), 0.15)$.
We initialised the\break  MCMC as follows. For $\sigma$ and $\phi$, we minimised
\[
\int_0^{25} \bigl(\hat{K}(r)^{0.25}-K(r;
\sigma, \phi)^{0.25}\bigr)^2 \,dr,
\]
where $K(r;\sigma, \phi)$ is the $K$-function of the model and $\hat
{K}(r)$ is Ripley's
estimate (Ripley \citeyear{Rip76}, \citeyear{Rip77}), resulting in initial values of $\sigma
=0.50$ and $\phi=12.66$. The initial value of $\Gamma$ was set to a
$256\times256$ matrix of zeros and $\beta$ was initialised using
estimates from an overdispersed Poisson generalised linear model fitted
to the cell counts, ignoring spatial correlation.

For the MCMC, we used a burn-in of 100,000 iterations followed by a
further 900,000 iterations,
of which we retained every 900th iteration so as to give a weakly dependent
sample of size 1000. Convergence and mixing diagnostics are shown in
the supplementary material [\citet{supp}].
Figure~\ref{fig:hickory_estimates} compares the prior and posterior
distributions of the
three model parameters showing, in particular, that the data give only
weak information
about the correlation range parameter, $\phi$. This is well known in
the classical geostatistical
context where the data are measured values of $S(x)$ (see, e.g.,
\cite*{Zha04}),
and is exacerbated in the point process setting.\looseness=1

The left plot in Figure~\ref{fig:hickory_surfaces} shows the pointwise
50th percentiles of the predictive distribution for the target, $\Lambda
(x)$ over the observation window; this clearly identifies the pattern
of the spatial variation in the intensity. The LGCP-based solution also
enables us to map areas of particularly low or high intensity. The
middle and right plots in Figure~\ref{fig:hickory_surfaces} are maps of
$\P\{\exp[S(x)]<1/2\}$ and $\P\{\exp[S(x)]>2\}$. The areas in these
plots where the posterior probabilities are high correspond,
respectively, to areas where the density of trees is less than half and
more than double the mean density.

%
%
%
%
%
%
%

The LGCP-based solution to the smoothing problem is arguably
over-elaborate by
comparison with simpler methods such as kernel smoothing. Against this,
arguments in
its favour are that it provides a principled rather than an \textit{ad
hoc} solution, probabilistic
prediction rather than point prediction, and an obvious extension to smoothing
in the presence of explanatory variables by specifying
$\Lambda(x) =\break  \exp\{u(x)^\prime\beta+ S(x)\}$, where $u(x)$ is a
vector of spatially referenced
explanatory variables.

\subsection{Spatial Segregation: Genotypic Diversity of Bovine
Tuberculosis in Cornwall, UK}
\label{sec:BTB}

Our second application concerns a multivariate version of the smoothing
problem described
in Section~\ref{sec:smoothing}. Events are now of $k$ types, hence, the
data are
$X = \{X_j\dvtx j=1,\ldots,k\}$, where
$X_j = \{x_{ij} \in A\dvtx i=1,\ldots,n_j\}$ and the corresponding intensity
functions are
$\lambda_j(x)\dvtx j=1,\ldots,k$. Write $\lambda(x) = \sum_{j=1}^k \lambda
_j(x)$ for the
intensity of the superposition. Under the additional assumption that
the underlying
process is an inhomogeneous Poisson process, then conditional on the
superposition,
the labellings of the events are a sequence of independent multinomial trials
with position-dependent multinomial probabilities,
\begin{eqnarray}
p_j(x) &= &\lambda_j(x)/\lambda(x) \nonumber\\
&=& \mathrm{P}(\mbox{event
at location } x \mbox{ is of type } j)\nonumber\\
 \eqntext{j=1,\ldots,k.}
\end{eqnarray}

A basic question for any multivariate point process data is whether the
type-specific component
processes are independent. When they are not, further questions of interest
are context-specific. Here, we describe an analysis of data relating to
bovine tuberculosis
in the county of Cornwall, UK.

\begin{figure}

\includegraphics{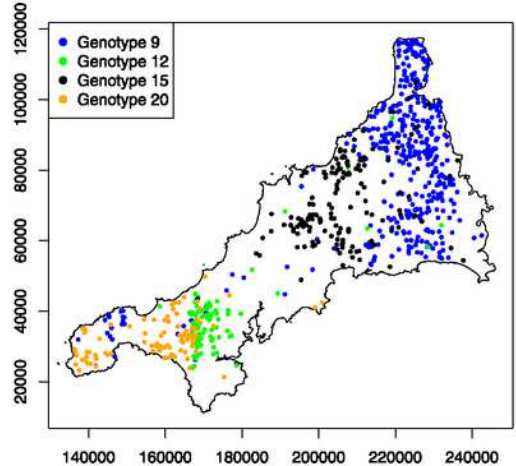}

\caption{Locations of cattle herds in Cornwall, UK, that have tested
positive for bovine tuberculosis (BTB) over
the period 1989 to 2002. Points are coded according to the genotype
of the infecting BTB organism.}
\label{fig:BTB_data}
\end{figure}

Bovine tuberculosis (BTB) is a serious disease of cattle. It is endemic
in parts of the UK. As part of the national control strategy, herds are
regularly inspected for BTB.
When disease in a herd is detected and at least one tuberculosis
bacterium is successfully cultured, the genotype that is responsible
for the BTB breakdown can be determined.
Here,
we re-visit an example from \citet{DigZheDur05} in which the
events are the
locations of cattle herds in the county of
Cornwall, UK, that have tested positive for bovine BTB over the period
1989 to 2002,
labelled according to their genotypes. The data, shown in Figure~\ref{fig:BTB_data},
are limited to the 873 locations with the four most
common genotypes; six less common genotypes accounted for an additional
46 cases.

The question of primary interest in this example is whether the
genotypes are randomly
intermingled amongst the locations and, if not, to what extent specific
genotypes are spatially
segregated. This question is of interest because the former would be
consistent with the major
transmission mechanism being cross-infection during the county-wide
movement of animals to
and from markets, whereas the latter would be indicative of local pools
of infection,
possibly involving\vadjust{\goodbreak} transmission between cattle and reservoirs of
infection in
local wildlife populations (Woodroffe {et al.}, \citeyear{Wooetal05}; \cite*{Donetal06}).

To model the data, we consider a multivariate log-Gaussian Cox process with
%
%
\begin{eqnarray}\label{eq:genotype_model}
\Lambda_k(x)=\exp\bigl(\beta_k + S_0(x) +
S_k(x)\bigr)
\nonumber
\\[-8pt]
\\[-8pt]
\eqntext{k=1,\ldots,m.}
\end{eqnarray}
In (\ref{eq:genotype_model}), $m=4$ is the number of genotypes, the parameters
$\beta_k$ relate to the intensities of the component processes,
$S_0(x)$ is a Gaussian process common to all types of points and the
$S_k(x)\dvtx k=1,\ldots,m$ are Gaussian processes specific to each genotype.
Although $S_0(x)$
is not identifiable from our data without additional assumptions,
its inclusion helps the interpretation of the model, in particular, by
emphasising
that the component intensities $\Lambda_k(x)$ are not mutually
independent processes.

\begin{figure*}

\includegraphics{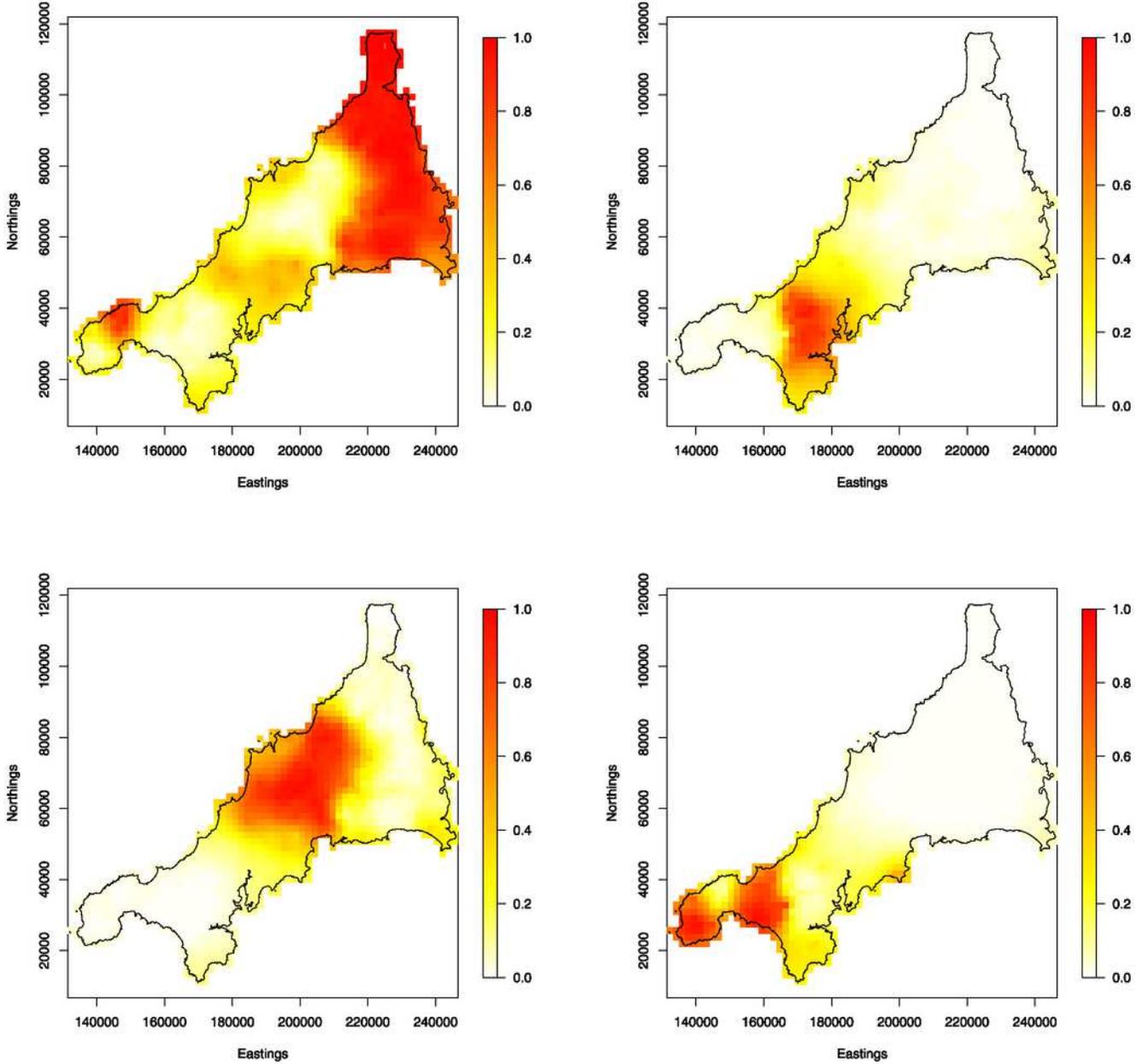}

\caption{Genotype-specific probability surfaces for the Cornwall BTB
data. Upper-left panel
corresponds to genotype 9, upper-right to genotype 12, lower-left to
genotype 15,
lower-right to genotype 20.}
\label{fig:type-specific}
\end{figure*}

In this example, we used informative priors for the model parameters:
$\log\sigma\sim\mathrm{N}(\log1.5,0.015)$,
$\log\phi\sim\mathrm{N}(\log 15\mbox{,}000,0.015)$ and $\beta_k \sim\mathrm{N}(0,10^6)$.
Because the algorithm mixes slowly, this proved to be a very challenging
computational problem.
For the MCMC, we used a burn-in of 100,000 iterations followed by a
further 18,000,000 iterations, of which we retained every 18,000th
iteration so as to give a sample of size 1000.
Convergence, mixing diagnostics and plots of the prior and posterior
distributions of $\sigma$ and $\phi$ are shown in the supplementary material [\citet{supp}].
These plots show that the chain appeared to have
reached stationarity with low autocorrelation in the thinned output.
The plots also illustrate that there is little information in the data
on $\sigma$ and~$\phi$.\vadjust{\goodbreak}

Within
(\ref{eq:genotype_model})
the hypothesis of randomly
intermingled genotypes corresponds to $S_k(x)=0\dvtx k=1,\ldots,4$, for all $x$.
Were it the case that
farms were uniformly distributed over Cornwall, $S_0(x)$ would then
represent the spatial
variation in the overall risk of BTB, irrespective of genotype.
Otherwise, $S_0(x)$
conflates spatial variation in overall risk with the spatial
distribution of farms. For the Cornwall BTB data
the evidence against randomly intermingled genotypes is overwhelming
and we focus our
attention on spatial variation in the probability that a case at
location $x$ is of type $k$,
for each of $k=1,\ldots,4$. These conditional probabilities
are
\[
p_k(x)=\frac{\Lambda_{k}(x)}{\sum_{j=1}^m \Lambda_{j}(x)} = \exp\biggl[-\sum
_{j \neq k} \bigl\{\beta_j + S_j(x)\bigr\}
\biggr]
\]
and do not depend on the unidentifiable common component $S_0(x)$.
Figure~\ref{fig:type-specific} shows point predictions of the
four genotype-specific probability
surfaces, defined as the conditional expectations $\hat{p}_k(x)=\break  \mathrm{E}[p_k(x)|X]$
for each of $k=1,\ldots,4$.

\begin{figure}

\includegraphics{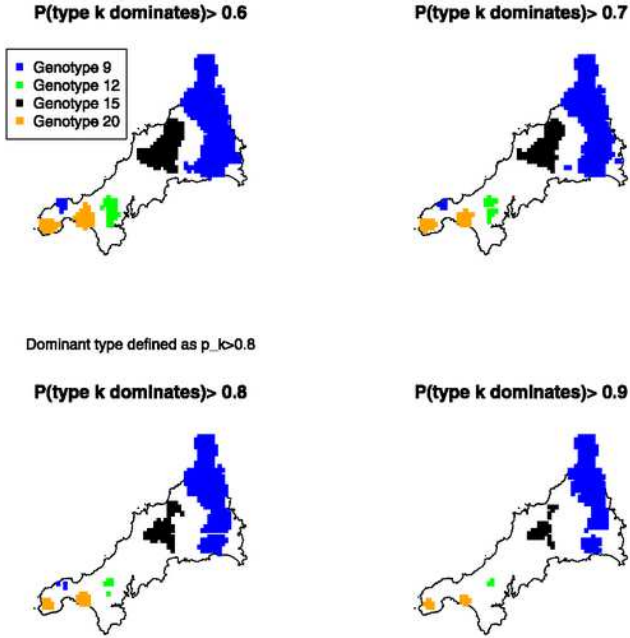}

\caption{$k$-dominant areas for each of the four genotypes in the
Cornwall data.}
\label{fig:segregation}
\end{figure}

As argued earlier, one advantage of a model-based approach to spatial
smoothing is that
results can be presented in ways that acknowledge the uncertainty on
the point predictions.
We could replace each panel of
Figure~\ref{fig:type-specific} by a set of percentile plots, as in
Figure~\ref{fig:hickory_surfaces}. For
an alternative display that focuses more directly
on the core issue of spatial segregation,
let $A_k(c,q)$ denote
the set of locations $x$ for which
$\mathrm{P}\{p_k(x)>c|X\}>q$. As $c$ and $q$ both approach
1, each $A_k(c,p)$ shrinks towards the empty set, but more slowly in a
highly segregated pattern
than in a weakly segregated one. In Figure~\ref{fig:segregation} we
show the areas
$A_k(0.8,q)$ for each of $q=0.6, 0.7, 0.8$ and 0.9. Genotype 9, which
contributes 494 to the
total of 873 cases, dominates strongly in an area to the east and less
strongly in a smaller area to the west. Genotype 15 contributes 166
cases and dominates in a
single, central area. Genotypes 12 and 20 each contribute a proportion of
approximately 0.12
to the total, with only small pockets of dominance to the south-west.

If infection times were known, we could perform inference via MCMC under
a spatio-temporal version of the model,
\begin{eqnarray}
\Lambda_k(x,t)=\exp\bigl(Z_k(x,t)\beta_k +
S_0(x,t) + S_k(x,t)\bigr)\nonumber\\
 \eqntext{k=1,\ldots,m,}
\end{eqnarray}
with $\Lambda_k(x,t)$, and $S_k(x,t)$ for $k=0,\ldots,m$
spatio-temporal versions of the purely spatial processes in (\ref
{eq:genotype_model})
and $Z_k(x,t)$ a vector of spatio-temporal covariates. Unlike purely spatial
models, spatio-temporal models are
potentially able to investigate mechanistic hypotheses about disease
transmission. For
example, in the context of this example a spatio-temporal analysis
could distinguish between
segregated patches that are stable over time or that grow from
initially isolated cases.

\subsection{Disease Atlases} \label{sect:disease_atlas}

\begin{figure*}

\includegraphics{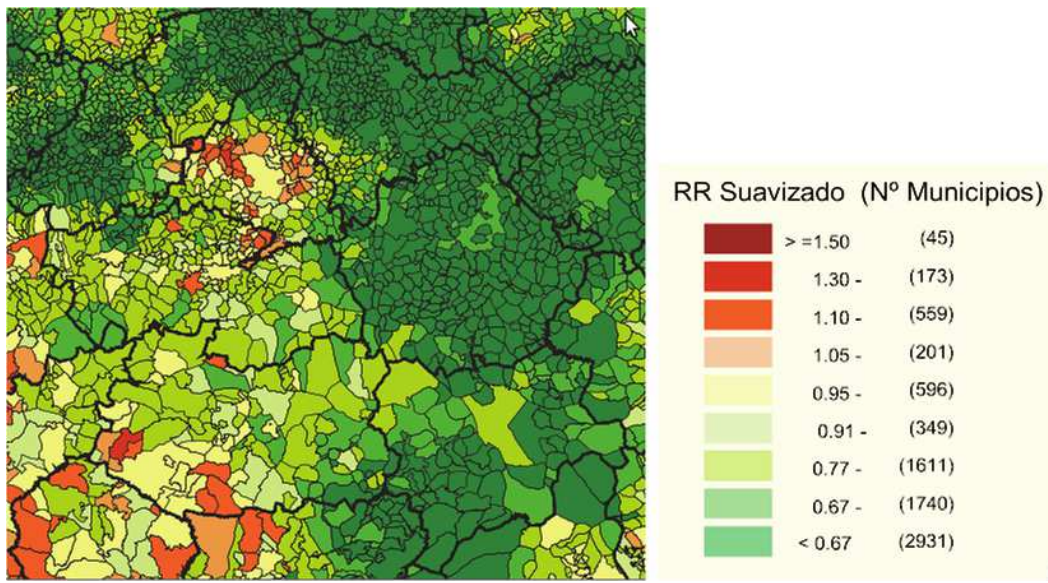}

\caption{Lung Cancer mortality in the Castile-La Mancha Region of
Spain. Figure reproduced from page 42 of L\'{o}pez-Abente {et al.} (\citeyear{Lopetal})
by kind permission of the authors.}
\label{fig:cancer_map1}
\end{figure*}

Figure~\ref{fig:cancer_map1}
is a typical example of the kind of map that appears in a variety of
cancer atlases. This example
is
taken from a Spanish national disease atlas project
(L\'{o}pez-Abente {et al.}, \citeyear{Lopetal}). The map estimates the spatial
variation in the
relative risk of lung cancer
in the Castile-La Mancha Region of Spain and some surrounding areas.
It is of a type known to geographers as a
choropleth map, in which the geographical region of interest,
$A$, is partitioned into
a set of subregions $A_i$ and each subregion is colour-coded according
to the numerical value
of the quantity of interest. The standard statistical methodology used
to convert data on
case-counts and the number of people at risk in each subregion is the following
hierarchical Poisson-Gaussian Markov random field model, due
to Besag, York and Moli\'{e} (\citeyear{BesYorMol91}).

Let $Y_i$ denote the number of cases in subregion $A_i$ and $E_i$
a standardised expectation computed as
the expected number of cases, taking into
account the demographics of the population in subregion $A_i$ but
assuming that risk is
otherwise spatially homogeneous. Assume that the $Y_i$ are
conditionally independent
Poisson-distributed conditional on a latent random vector $S =
(S_1,\ldots,S_m)$, with
conditional means
$\mu_i = E_i \exp(\alpha+S_i)$. Finally, assume that $S$ is
multivariate Gaussian, with
its distribution specified as a Gaussian Markov random field (\cite{RueHel05}).
A Markov random field is a multivariate distribution specified indirectly
by its full conditionals, $[S_i|S_j\dvtx j \neq i]$. In the
Besag, York and Moli\'{e} (\citeyear{BesYorMol91}) model the full
conditionals take the so-called \textit{intrinsic autoregressive} form,
%
%
\begin{equation}
S_i|S_j\dvtx j \neq i \sim\mathrm{N}\bigl(
\bar{S}_i,\tau^2/n_i\bigr),
\label{eq:intrinsic}
\end{equation}
where $\bar{S}_i = n_i^{-1} \sum_{j \sim i} S_j$ is the mean of the $S_j$
over subregions $A_j$ considered to be \textit{neighbours} of $A_i$ and
$n_i$ is the
number of such neighbours. Typically, subregions are defined to be
neighbours if they
share a common boundary.

An alternative approach is to model the locations of individual cancer
cases as an LGCP
with
intensity $\Lambda(x) = d(x) R(x)$, where $d(x)$ represents population
density, assumed
known, and $R(x)$ denotes
disease risk, $R(x) = \exp\{S(x)\}$. Conditional on $R(\cdot)$,
case-counts in subregions
$A_i$ are independent and Poisson-distributed with means
\[
\mu_i = \int_{A_i} \,d(x) R(x) \,dx.
\]
This approach leads to spatially smooth risk-maps whose interpretation
is independent of the particular partition of $A$ into subregions
$A_i$. This is
an important
consideration when the $A_i$ differ greatly in size and shape,
as the definition of neighbours
in an MRF model then becomes problematic; see, for example, \citet{Wal04}.
Fitting a spatially continuous model also has the potential to add information
to an analysis of aggregated data, for example, when data on
environmental risk-factors are available at high spatial resolution.
A caveat is that the population density may only be available
in the form of small-area population counts,
implying
a piece-wise constant surface $d(x)$ that can only be a convenient
fiction. Note, however,
that spatially continuous modelled population density maps have been
constructed and are
freely available; see, for\break  example,
\href{http://sedac.ciesin.columbia.edu/data/set/gpw-v3-population-density}{http://sedac.ciesin.columbia.edu/data/}\break
\href{http://sedac.ciesin.columbia.edu/data/set/gpw-v3-population-density}{set/gpw-v3-population-density}.

For the Spanish lung cancer data, we have covariate information
available at small-area,\vadjust{\goodbreak}
which we incorporate by fitting the model
%
%
\begin{equation}
\Lambda(x) = d(x) \exp\bigl\{z(x)^\prime\beta+ S(x)\bigr\},
\label{eqn:aggregate1}
\end{equation}
treating the covariate surfaces $z(x)$ as piece-wise constant.

For Bayesian inference under the continuous model (\ref
{eqn:aggregate1}) we follow
Li {et al.} (\citeyear{Lietal12})
by adding standard data augmentation techniques to the MCMC fitting algorithm
described earlier.
Recall that for computational purposes, we perform all calculations
on a fine grid, treating the cell counts in each grid cell as Poisson
distributed
conditional on the latent process $S(\cdot)$.
Provided the computational grid is fine enough, each
$A_i$ can be approximated by the union of
a set of grid cells, and we can use a grid-based
Gibbs sampling strategy, repeatedly sampling first from
$[S,\beta,\theta|N,Y_{+}] = [S,\beta,\theta|N]$ and then from
$[N|S,\beta,\theta,Y_{+}]$,
where $N$ are the cell counts on the computational grid, $Y_{+}=\{
Y_i=\sum_{x\in A_i}N(x)\dvtx i=\break 1,\ldots,m\}$ and $\theta$ parameterises the
covariance structure of $S$.
Sampling from the first of these densities can be achieved using a
Metropolis-Hastings update as discussed in Section~\ref{sect:computation}.
 The second density is a multinomial distribution
and poses no difficulty.


Our priors for this example were as follows: $\log\sigma\sim\mathrm{N}(\log1,0.3)$,
$\log\phi\sim\mathrm{N}(\log3000,0.15)$ and
$\beta\sim\break\operatorname{MVN}(0,10^6 I)$.
For the MCMC algorithm, we used a burn-in of 100,000 iterations
followed by a further 18,000,000 iterations, of which we retained every
18,000th iteration so as to give a sample of size 1000. Convergence,
mixing diagnostics and plots of the prior and posterior distributions
of $\sigma$ and $\phi$ are shown in the supplementary material [\citet{supp}]. As in
the Cornwall BTB analysis,
these plots indicated convergence to
the stationary distribution and low autocorrelation in the thinned output.

\begin{figure}

\includegraphics{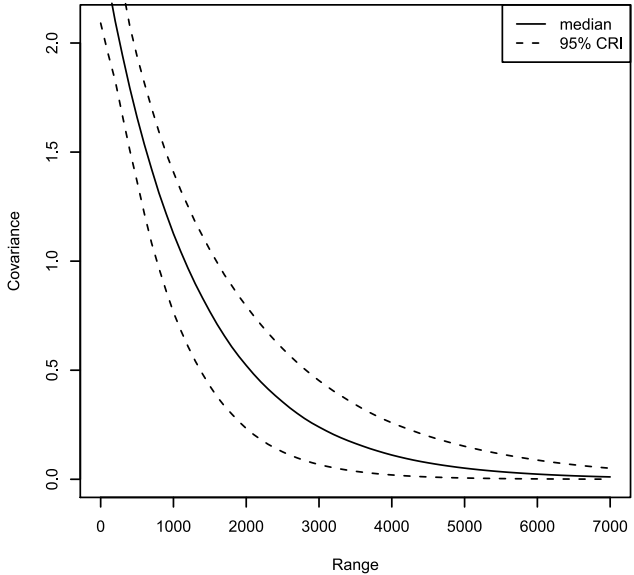}

\caption{Posterior covariance function.}
\label{fig:lung_cov_fct}
\end{figure}

In the analysis reported here, we base our offset on modelled
population data at 100 metre resolution obtained from the European
Environment Agen\-cy; see
\href{http://www.eea.europa.eu/data-and-maps/data/population-density-disaggregated-with-corine-land-cover-2000-2}{http://www.eea.europa.eu/data-and-maps/}\break
\href{http://www.eea.europa.eu/data-and-maps/data/population-density-disaggregated-with-corine-land-cover-2000-2}{data/population-density-disaggregated-with-corine-}\break
\href{http://www.eea.europa.eu/data-and-maps/data/population-density-disaggregated-with-corine-land-cover-2000-2}{land-cover-2000-2}.
We projected this very fine population information onto our
computational grid, which consisted of cells $3100\times3100$ metres in
dimension. We used an exponential model for the covariance function of
$S(\cdot)$ and estimated its
parameters (posterior median and 95\% credible interval) to be $\sigma
=1.57\ (1.45,1.71)$ and $\phi=1294\ (814,1849)$ metres. Figure~\ref{fig:lung_cov_fct} illustrates the shape of the posterior covariance
function; it can be seen from this plot that the posterior dependence
between cells is over a relatively small range.

Table~\ref{tab:coveffects_lung} summarises our estimation of covariate
effects. Our results show that estimated (posterior median) mortality
rates were higher in areas with higher rates of illiteracy and higher
income; these effects were statistically significant at the 5\% level,
in the sense
that the Bayesian 95\% credible intervals excluded zero.
The remaining covariates (unemployment, percentage farmers, percentage
of people over 65 and average number of people per home) had a
protective effect, but only significantly so in the
case of
percentage farmers.

\begin{table}
\caption{Selected quantiles of the posterior distributions of
standardised covariate effects for the Spanish lung cancer~data}\label{tab:coveffects_lung}
\begin{tabular*}{\columnwidth}{@{\extracolsep{\fill}}lccc@{}}
\hline
& \multicolumn{3}{c@{}}{\textbf{Quantile}} \\[-6pt]
& \multicolumn{3}{c@{}}{\hrulefill} \\
\textbf{Parameter} & \textbf{0.50} & \textbf{0.025} & \textbf{0.975} \\
\hline
Percentage illiterate & 1.13 & 1.03 & 1.24 \\
Percentage unemployed & 0.92 & 0.8\phantom{0} & 1.03 \\
Percentage farmers & 0.88 & 0.76 & 1.00 \\
Percentage of people over 65 years old & 1.2\phantom{0} & 0.96 & 1.51 \\
Income index & 1.19 & 1.03 & 1.39 \\
Average number of people per home & 0.98 & 0.75 & 1.26 \\
\hline
\end{tabular*}
\end{table}

\begin{figure*}

\includegraphics{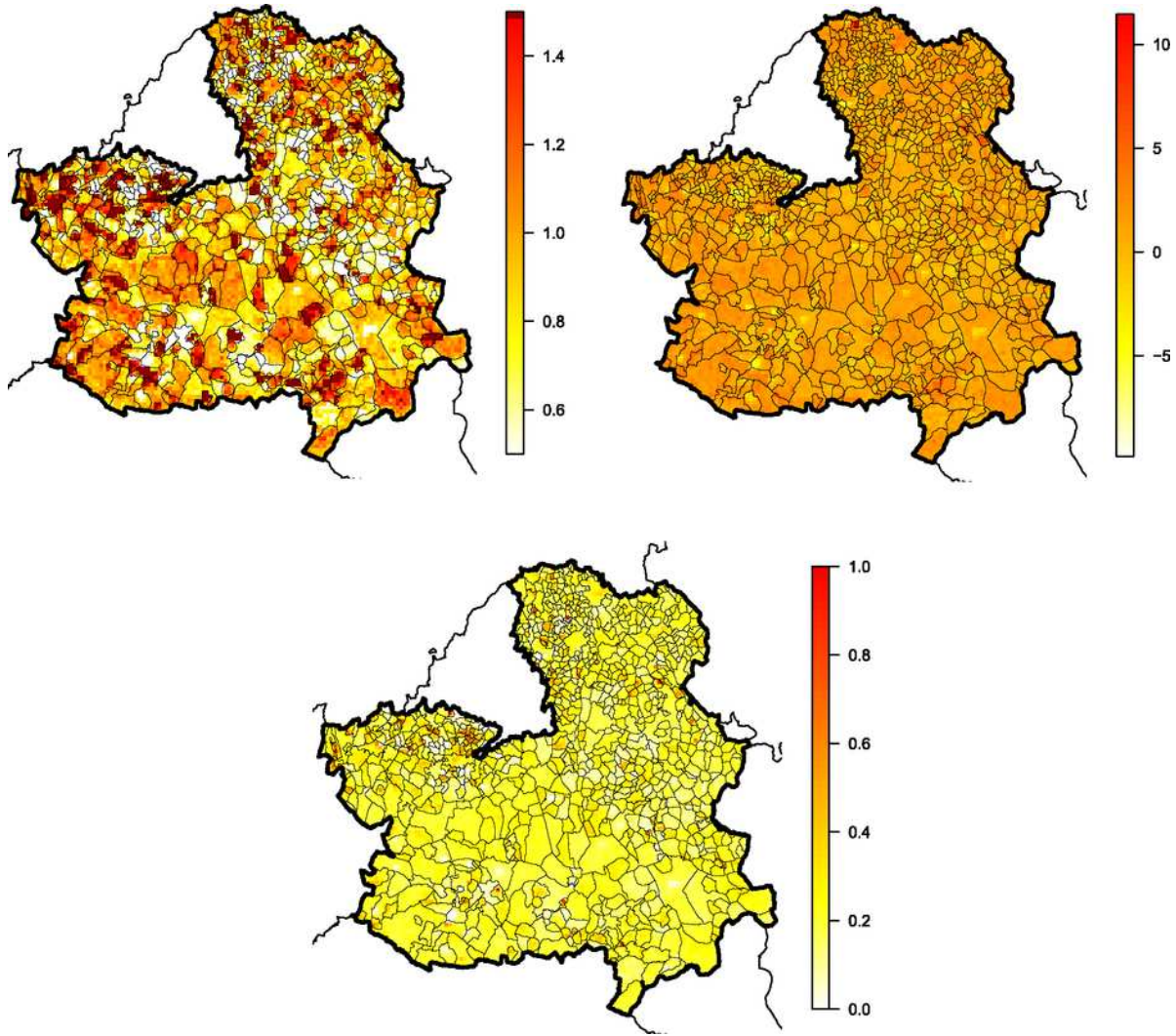}

\caption{Lung Cancer mortality in the Castile-La Mancha Region of Spain.
The top left panel shows covariate-adjusted relative risk. The range of
values was restricted to lie between
0.5 and 1.5 to allow comparison with Figure \protect\ref{fig:cancer_map1}.
Inside
the Castile-La Mancha region, cells with mean relative risk greater
than 1.5 appear
dark red and cells with relative risk below 0.5 appear white. The top
right panel shows the log of the estimated variance of relative risk.
The bottom
panel shows the predictive probability that the covariate-adjusted
relative risk exceeds~1.1.}
\label{fig:lung_lgcp}
\end{figure*}

Figure~\ref{fig:lung_lgcp} shows the resulting maps. The top left-hand
panel shows the predicted, covariate-adjusted relative risk surface
derived from the log-Gaussian Cox process model (\ref{eqn:aggregate1}).
This predicted relative risk surface
reveals several small areas of raised risk that are not apparent in
Figure~\ref{fig:cancer_map1}. The top right-hand panel shows the log of
the estimated variance of relative risk. To account for this variation,
we produced a plot of the posterior probability that relative risk
exceeds 1.1, shown in the bottom panel. This shows that higher rates of
incidence appear to be mainly confined to a number of small townships,
the largest of which is an area to the
north of Toledo and surrounding the Illescas municipality, where there
are a number of contiguous cells for which the probability exceeds 0.6.

We acknowledge that this is an illustrative example. In particular,
we cannot guarantee the reliability of the estimate of population density
used as an offset.

In a discussion of Markov models for spatial data, \citet{Wal04}
investigated properties of the covariance structure implied by the
simultaneous and conditional autoregressive models on an irregular
lattice. She concluded that the ``implied spatial correlation [between
cells in these] models does not seem to follow an intuitive or
practical scheme'' and advises ``[using] other ways of modelling lattice
data $\ldots$ should be considered, especially when there is interest
in understanding the spatial structure''. Our approach is one such. Others,
which we discuss in Section~\ref{sec:synthesis}, include proposals in
Best, Ickstadt and Wolpert (\citeyear{BesIckWol00}) and \citet{KelWak02}.

Our spatially continuous formulation does not
entirely rescue us from the trap of the ecological fallacy (Piantadosi,
Byar and Green, \citeyear{PiaByaGre88};
\cite*{GreMor90}).
In a spatial context, this
refers to the fact that the association between a risk-factor and a
health outcome need not be,
and usually is not, independent of the spatial scale on which the
risk-factor and outcome
variables are defined. In our example, we have to accept that
treating covariate surfaces as if they were piece-wise constant is a
convenient fiction. However,
our methodology avoids any necessity to aggregate all covariate and
outcome variables to
a common set of spatial units, but rather operates at the fine
resolution of the computational
grid.
In effect, this
enables us to place a spatially continuous interpretation
on any parameters relating to continuously measured components of the
model, whether covariates
or the latent stochastic process $S(x)$.

\section{Spatio-Temporal Log-Gaussian Cox Processes}\label{sec6}

\subsection{Models}\label{sec6.1}

A spatio-temporal LGCP is defined in the obvious way, as a spatio-temporal
Poisson point process conditional on the realisation of a stochastic
intensity function
$\Lambda(x,t) = \exp\{S(x,t)\}$, where $S(\cdot)$ is a Gaussian process.
\citet{GneGut10N2} review the
literature on formulating models for spatio-temporal Gaussian
processes. They make a useful
distinction between physically motivated constructions and more
empirical formulations.
An example
of the former is given in Brown {et al.} (\citeyear{Broetal00}), who propose models
based on a physical dispersion
process. In discrete time, with $\delta$ denoting the time-separation
between successive
realisations of the spatial field, their model takes the form
%
%
\begin{eqnarray}\label{eq:blur}
&& S(x,t)
\nonumber
\\[-8pt]
\\[-8pt]
\nonumber
&&\quad = \int h_\delta(u) S(x-u,t-\delta) \,du + Z_\delta(x,t),
\end{eqnarray}
where $h_\delta(\cdot)$ is a smoothing kernel and $Z_\delta(\cdot)$ is
a noise process,
in each case with parameters that depend on the value of $\delta$ in
such a way as to
give a consistent interpretation in the spatio-temporally continuous
limit
as $\delta\rightarrow0$.

Amongst empirical spatio-temporal covariance\break  models, a basic
distinction is between separable
and nonseparable models. Suppose that $S(x,t)$ is stationary, with
variance $\sigma^2$
and correlation function $r(u,v) = \operatorname{Corr}\{S(x,t),S(x-u,t-v)\}$. In
a \textit{separable}
model, $r(u,v) = r_1(u) r_2(v)$, where $r_1(\cdot)$ and $r_2(\cdot)$ are
spatial and temporal correlation functions. The separability assumption
is convenient, not
least because any valid specification of $r_1(u)$ and $r_2(v)$
guarantees the
validity of $r(u,v)$, but it is not especially natural. Parametric
families of
nonseparable models are discussed in \citet{CreHua99}, \citet{Gne02}, Ma (\citeyear{Ma03}, \citeyear{Ma08})
and \citet{RodDig10}.

As noted by \citet{GneGut10N2}, whilst spatio-temporally continuous
processes are, in formal mathematical terms, simply spatially
continuous processes with an extra dimension, from a
scientific perspective models need to reflect the fundamentally
different nature of
space and time, and, in particular, time's directional quality. For
this reason,
in applications where data arise as a set of spatially indexed
time-series, a
natural way to formulate a spatio-temporal model is as a multivariate
time series whose cross-covariance functions are spatially structured.
For example,
a spatially discrete version of (\ref{eq:blur}) on a finite set of spatial
locations $x_i\dvtx i=1,\ldots,n$ and integer times $t$ would be
%
%
\begin{equation}
S_{it} = \sum_{j=1}^n
h_{ij} S_{i,t-1} + Z_{it}, \label{eq:discrete_blur}
\end{equation}
where the $h_{ij}$ are functions of the corresponding locations, $x_i$
and $x_j$.
For a review of models of this kind, see
\citet{Gam10}.

\subsection{Spatio-Temporal Prediction: Real-Time Monitoring of
Gastrointestinal Disease}\label{sec6.2}

An early implementation of spatio-temporal log-Gaussian process
modelling was used in the AEGISS project (Ascertainment and Enhancement
of\break
Gastroenteric Infection Surveillance Statistics, see\break
\href{http://www.maths.lancs.ac.uk/\textasciitilde diggle/Aegiss/day.html\%3fyear=2002}{http://www.maths.lancs.ac.uk/\textasciitilde diggle/Aegiss/day.}\break
\href{http://www.maths.lancs.ac.uk/\textasciitilde diggle/Aegiss/day.html\%3fyear=2002}{html\%3fyear=2002}).
The overall aim of the project was to investigate how
health-care data routinely collected within the UK's National Health
Service (NHS)
could be used to spot outbreaks of gastro-intestinal disease. The
project is
described in detail in Diggle {et al.} (\citeyear{Dietal03}),
whilst \citet{DigRowSu05} give details of the
spatio-temporal statistical model.

As part of the government's modernisation\break  programme for the NHS, the
nonemergency NHS Direct telephone service was launched in the late
1990s, and by 2000 was serving all of England\break  and Wales
(\href{http://www.nhsdirect.nhs.uk/About/WhatIsNHSDirect/History}{http://www.nhsdirect.nhs.uk/About/}\break
\href{http://www.nhsdirect.nhs.uk/About/WhatIsNHSDirect/History}{WhatIsNHSDirect/History}).
Callers to
this 24-hour system were questioned about their problem and advised
accordingly. This process reduced calls to an ``algorithm code'' which
was a broad classification of the problem. Basic information on the
caller, including age, sex and postal code, was also recorded. \citet{CooChi04} give a more detailed description of the NHS Direct
system. \citet{MarShe04} analyse its impact on the demand for
primary care in the UK. Cooper {et al.} (\citeyear{Cooetal03}) report a
retrospective analysis
of 150,000 calls to NHS Direct
classified as diarrhoea or vomiting, and
concluded that fluctuations in the rate of such calls could be a useful proxy
for monitoring the incidence of gastrointestinal illness.

In the AEGISS project, residential
postal codes associated with calls classified as relating
to diarrhoea or vomiting
were converted to grid references using a
lookup table. Postal codes at this level are referenced to 100 metre
precision, which on the scale of the study area (the county of
Hampshire) is effectively continuous. The data then formed a
spatio-temporal point pattern.

\begin{figure*}

\includegraphics{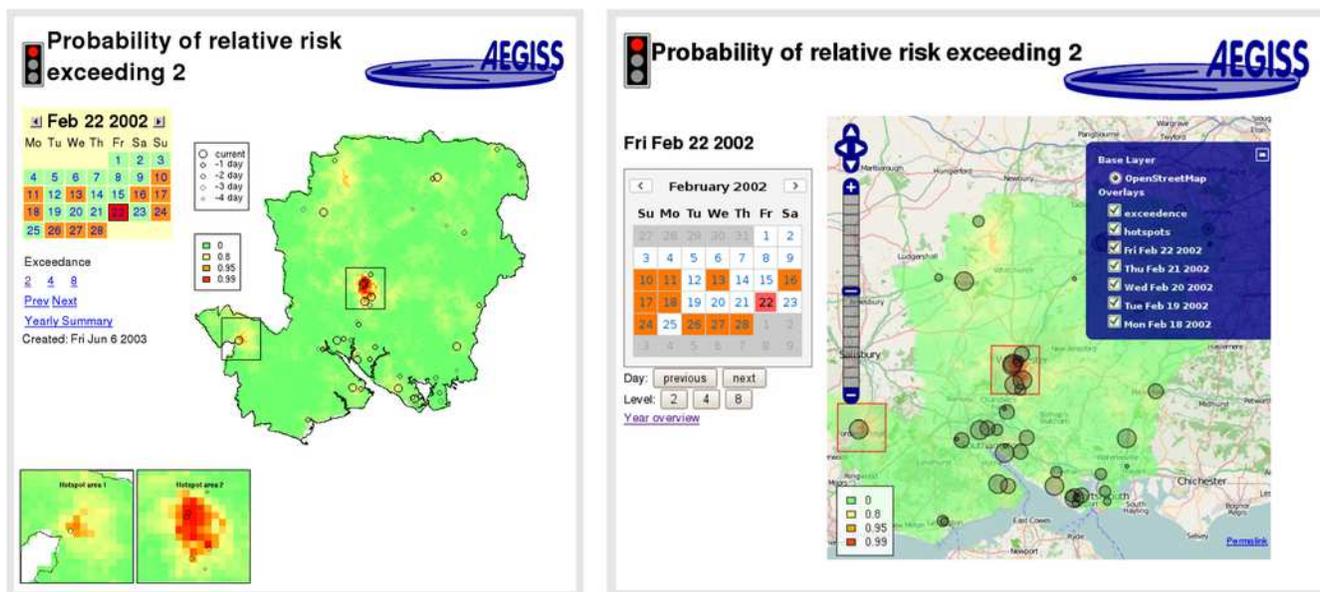}

\caption{AEGISS web page design. Left-hand panel shows the original
design, right-hand panel
a modern redesign.}
\label{fig:aegiss1}
\end{figure*}

The daily extraction of data for Hampshire and the location coding was
done by the NHS at South\-ampton. These data were encrypted and sent
by email to Lancaster, where the emails were automatically filtered,
decrypted and stored. An overnight run of the MALA algorithm
described in \citet{BriDig01}
took the latest data and produced maps of predictive probabilities for
the risk exceeding multiples 2, 4 and 8 of the baseline rate.

The specification of the model, based on an exploratory analysis of
the data, was a spatio-temporal LGCP with intensity
\[
\Lambda(x,t) = \lambda_0(x)\mu_0(t)\exp\bigl\{S(x,t)
\bigr\}.
\]
The spatial baseline component, $\lambda_0(x)$, was calculated by a
kernel smoothing of the first two years of case locations, whilst the
temporal baseline, $\mu_0(t)$, was obtained by fitting a standard
Poisson regression model to the counts over time. This regression
model included an annual seasonal component, a factor representing the
day-of-the-week and a trend term to represent the increasing take-up
of the NHS Direct service during the life-time of the project.

The parameters of $S(x,t)$ were then estimated using moment-based
methods, as in \citet{BriDig01},
with a separable correlation structure. Uncertainty
in these parameter estimates was considered to
have a minimal effect on the
predictive distribution of $S(x,t)$ because parameter estimates are
informed by
all of the data, whereas prediction of $S(x,t)$ given the model
parameters benefits only from data points that lie close to $(x,t)$,
that is, within the range of
the spatio-temporal correlation.\looseness=1

Plug-in predictive inference was then performed using the MALA algorithm
on each new set of data
arriving overnight. Instead of storing the outputs from each of 10,000
iterations, only a count of where $S(x,t)$ exceeded a threshold
that corresponded to 2, 4 or 8 times the baseline risk
was retained. This range of thresholds was chosen in consultation with
clinicians; a doubling of risk was considered of possible interest,
whilst an
eightfold increase was considered potentially serious.
These exceedence counts
were then converted into exceedence probabilities.

Presentation of these exceedence maps was an important aspect of the
AE\-GISS project. At the time, there were few implementations of maps on
the inter\-net---UMN MapServer was released as open source in 1997 and
the Google Maps service started in 2005.
A simpler approach was used where static images of the exceedence
probabilities were generated by \texttt{R}'s graphics system. Regions
where the
exceedence probability was higher than 0.9 were outlined with a box
and displayed in a zoomed-in version below the main
graphic. Other page controls enabled the user to select the threshold
value as 2, 4 or 8, and to select a day or month. A traffic light
system of green, amber and red warnings dependent on the severity of
exceedence threshold crossings was developed for rapid assessment of
conditions on any particular day. The left-hand panel of
Figure~\ref{fig:aegiss1} shows a day
where two clusters of grid cells show high predictive probability of
at least a doubling of risk relative to baseline.

With modern web-based technologies the user interface could be constructed
as a dynamic web-\break mapping system that would allow the user freely to navigate
the study region. Layers of information, such as cases or exceedence
probability maps, can then be selected by the user as overlays.
The right-hand panel of Figure~\ref{fig:aegiss1} shows the same day as the left-hand panel, but uses
the OpenLayers (\href{http://www.openlayers.org}{http://}
\href{http://www.openlayers.org}{www.openlayers.org}) web-mapping
toolkit to superimpose the
cases and risk surface on a base map composed of data from
OpenStreetMap (\href{http://www.openstreetmap.org}{http://}\break
\href{http://www.openstreetmap.org}{www.openstreetmap.org}). This also shows the layer
selector menu for further customisation.

Increases in computing power and algorithmic advances mean that longer
MCMC runs can be performed overnight or on finer spatial
resolutions. However, increasing ethical concerns over data use and
patient confidentiality mean that finely resolved\break  spatio-temporal data
are becoming harder to obtain. Recent changes in the organisation of
the NHS
24-hour telephone helpline has meant that several providers will now be
responsible for regional services contributing to a new system,
NHS111 (\href{http://www.nhs.uk/111}{http://}
\href{http://www.nhs.uk/111}{www.nhs.uk/111}).
AEGISS was originally conceived as a pilot project that could
be rolled out to all of the UK, but obtaining data from all the new
providers and dealing with possible systematic differences between
them in order to perform a statistically rigorous analysis is now more
challenging. The future of health surveillance systems may lie in the
use of
multivariate
spatio-temporal models to combine information
from multiple data streams including nontraditional proxies
for health outcomes, such as nonprescription medicine sales, counts of
key words and phrases used in
search engine queries, and text-mining of social media sites.

\section{Data Synthesis: Integrated Analysis of Exposure and Health Outcome
Data at Multiple Spatial Scales} \label{sec:synthesis}

The ubiquitous problem of dealing with exposure and health outcome data recorded
at disparate spatial
scales is known to geographers as the ``modifiable areal unit
problem.'' See, for example, the reviews by
\citet{GotYou02} and \citet{DarBra07}. In the statistical
literature, a more common term is
``spatial misalignment.'' See, for example, \citet{Gel10}. Several
authors have considered special cases
of this problem in an epidemiological setting. Mugglin, Carlin and
Gelfand (\citeyear{MugCarGel00})
deal with data in the form of disease
counts on a partition of the region of interest, $A$, into a discrete
set of subregions,
$A_i$, together with covariate
information on a different partition,
$B_i$, say. Their solution is based on creating a single, finer partition
that includes all nonzero intersections $A_i \cap B_j$ . Best, Ickstadt
and Wolpert (\citeyear{BesIckWol00})
also consider count data on a discrete
partition of $A$, but assume that covariate information on a risk
factor of interest is available throughout $A$.
They consider count data to be
derived from an underlying Cox process whose intensity varies
in a spatially continuous manner through the combination of a covariate
effect and a latent stochastic
process modelled as a kernel-smoothed gamma random field. They then
derive the distribution of the
observed counts by spatial integration over the $A_i$. \citet{KelWak02}
take a similar approach,
but using a log-Gaussian latent stochastic process rather than a gamma
random field. The technical and
computational issues that arise when handling spatial integrals of
stochastic processes can be simplified
by using low-rank models, such as the class of Gaussian predictive
process models proposed by Banerjee {et al.} (\citeyear{Banetal08})
and further developed by Finley {et~al.} (\citeyear{Finetal09}). \citet{Gel12}
gives a useful summary of this
and related work.

All of these approaches can be subsumed within a single
modelling framework for multiple exposures and disease risk by
considering these
as a set of
spatially continuous processes, irrespective of the spatial resolution at
which data elements are recorded.
For example, a model for the spatial association between disease risk,
$R(x)$, and
$m$ exposures
$T_k(x) \dvtx k = 1,\ldots,m$ can be obtained by treating individual case-locations
as an LGCP with intensity
%
%
\begin{equation}
R(x) = \exp\Biggl\{\alpha+\sum_{k=1}^p
\beta_k T_k(x) + S(x)\Biggr\}, \label{eq:multicox}
\end{equation}
where $S(x)$ denotes stochastic variation in risk that is not captured
by the $p$ covariate
processes $T_k(x)$.
The inferential algorithms associated with model (\ref{eq:multicox})
would then depend on the
structure of the available data.

Suppose, for example, that
health outcome data are available in the form of area-level counts,
$Y_i\dvtx i=1,\ldots,n$,
in subregions $A_i$,
whilst exposure data are obtained as collections of unbiased estimates,
$U_{ik}$, of the
$T_k(x)$ at corresponding locations $x_{ik}\dvtx i=1,\ldots,m_k$. Suppose further
that the $U_{ik}$ are conditionally independent, with
$U_{ik} | T_k(\cdot) \sim\mathrm{N}(T_k(x_{ik}),\tau_k^2)$, the
processes $T_k(\cdot)$ are
jointly Gaussian and the process $S(\cdot)$ is also Gaussian and independent
of the $T_k(\cdot)$. A~possible inferential goal is to evaluate the
predictive distribution of
the risk surface $R(\cdot)$ given the data $Y_i\dvtx i=1,\ldots,m$ and
$U_{ik}\dvtx i=1,\ldots,m_k;k=1,\ldots,p$.
In an obvious shorthand, and temporarily ignoring the issue of parameter
estimation, the required predictive distribution is $[S,T|U,Y]$. The
joint distribution of
$S$, $T$, $U$ and $Y$ factorises as
%
%
\begin{equation}
[S,T,U,Y] = [S] [T] [U|T] [Y|S,T], \label{eq:factorise}
\end{equation}
where $[S]$ and $[T]$ are multivariate Gaussian densities, $[U|T]$ is a
product of
univariate Gaussian densities, and $[Y|S,T]$ is a product of Poisson probability
distributions with
means
\[
\mu_i = \int_{A_i} R(x) \,dx.
\]
Sampling from the required predictive distributions can then proceed
using a
suitable MCMC algorithm.
For Bayesian parameter estimation, we would
augment (\ref{eq:factorise}) by a suitable joint prior
for the model parameters before designing the MCMC algorithm.

A specific example of data synthesis
concerns an ongoing leptospirosis cohort study in a poor community within
the city of Salvador, Brazil.
Leptospirosis is considered to be the most widespread of the zoonotic diseases.
This is due to the large number of people worldwide, but especially in poor
communities, who live in close proximity to wild and domestic mammals
that serve as reservoirs of infection and shed the agent
in their urine. The major mode of transmission
is contact with contaminated water or soil (\cite{Lev01}; \cite{Bhaetal03};
McBride {et~al.}, \citeyear{McBetal05}).
In the majority of cases
infection leads to an
asymptomatic or mild,
self-limiting febrile illness.
However, severe cases can lead to potentially fatal
acute renal failure and pulmonary haemorrhage
syndrome. Leptospirosis
is traditionally associated with rural-based subsistence\break
farming communities,
but rapid urbanization and widening
social inequality have led to the dramatic growth of urban slums, where
the lack of basic sanitation favours
rat-borne transmission (Ko {et al.}, \citeyear{Koetal99}; Johnson {et al.}, \citeyear{Johetal04}).

The goals of the cohort
study are to investigate the combined effects of social and physical
environmental
factors on disease risk, and to map
the unexplained spatio-temporal variation in incidence.
In the study, approximately 1700 subjects $i=1,\ldots,n$ at residential
locations $x_i$
provide blood-samples on recruitment and at
subsequent times $t_{ij}$ approximately 6, 12, 18 and 24 months later.
At each post-recruitment visit,
sero-conversion is defined as a\break  change from zero to
positive, or
at least a fourfold increase in concentration.
The resulting data consist of
binary responses, $Y_{ij}=0/1\dvtx j= 1, 2, 3, 4$ (sero-conversion no/yes),
together with a mix of time-constant and time-varying risk-factors, $r_{ij}$.

A conventional analysis might treat the data from each subject as a
time-sequence
of binary responses with associated explanatory variables. Widely used methods
for data of this kind include generalised estimating equations (\cite{LiaZeg86})
and generalised linear mixed models (\cite{BreCla93}). An
analysis more
in keeping with the philosophy of the current paper would proceed as follows.

Let $a_i$ and $b_i(t)$ denote time-constant and time-varying
explanatory variables
associated with subject~$i$, and
$t_{ij}$ the times at which blood samples are
taken, setting $t_{i0}=0$ for all $i$. Note that explanatory variables
can be of
two distinct kinds: characteristics of an individual subject, for example,
their age; and
characteristics of a subject's place of residence, for example,
its proximity to an open-sewer. In principle, the latter can be indexed
by a spatially
continuous location, hence, $a_i=A(x_i)$ and $b_i(t) = B(x_i,t)$.
A response $Y_{ij}=1$ indicates that at least one infection event has
occurred in the
time-interval $(t_{i,j-1},t_{ij})$. A~model for each subject's risk of infection
then requires the specification of a set of person-specific hazard functions,
$\Lambda_i(t)$. A~model that allows for unmeasured risk factors would
be a
set of LGCPs, one for each subject, with respective stochastic intensities,
%
%
\begin{equation}\qquad
\Lambda_i(t) = \exp\bigl\{a_i^\prime\alpha+
b_i(t_{ij})^\prime\beta+ U_i +
S(x_i,t)\bigr\}, \label{eqn:hazard_model}
\end{equation}
where the $U_i$ are mutually independent $\mathrm{N}(0,\nu^2)$ and
$S(x,t)$ is a
spatio-temporally continuous
Gaussian process. It follows that
%
%
\begin{eqnarray}\label{eqn:hazard_cumulative}
&&\mathrm{P}\bigl\{Y_{it}=1 | \Lambda_i(\cdot)\bigr\}
\nonumber
\\[-8pt]
\\[-8pt]
\nonumber
&&\quad= 1 -
\exp \biggl\{-\int_{t_{i,j-1}}^{t_{ij}} \Lambda_i(u)
\,du \biggr\}.
\end{eqnarray}
In practice,
values of $a(x)$ and $b(x,t)$ may only
be observed incompletely, either at a finite number of locations or as
small-area averages.
For notational convenience, we consider only a single, incompletely observed
spatio-temporal covariate whose measured values, $b_k\dvtx k=1,\ldots,m$, we
model as
%
%
\begin{equation}
b_k = B(x_k,t_k) + Z_k,
\label{eqn:Bmodel}
\end{equation}
where $B(x,t)$ is a spatio-temporal Gaussian process and the $Z_k$ are
mutually independent
$\mathrm{N}(0,\tau^2)$ measurement errors. Then, (\ref{eqn:hazard_model}) becomes
%
%
\begin{equation}
\Lambda_i(t) = \exp\bigl\{B(x_{i},t_{ij})^\prime
\beta+ U_i + S(x_i,t)\bigr\} \label{eqn:hazard_model2}.
\end{equation}
Inference for the model defined by (\ref{eqn:hazard_cumulative}),
(\ref{eqn:Bmodel}) and (\ref{eqn:hazard_model2}), based on data
$\{y_{ij}\dvtx j=1,\ldots,4; i=1,\ldots,n\}$ and
$b=\{b_k\dvtx k=1,\ldots,m\}$,
would require further development of MCMC algorithms of the kind
de-\break scribed in
Section~\ref{sect:computation}.

\section{Discussion}\label{sec8}

In this paper we have argued that the LGCP provides a useful class of
models, not only
for point process data but also for any problem involving prediction of an
incompletely observed spatial or spatio-temporal process, irrespective
of data format. Developments in statistical computation have made
the combination of likelihood-based, classical or Bayesian
parameter estimation and probabilistic prediction
feasible for relatively large data sets, including real-time updating
of spatio-temporal
predictions.

In each of our applications, the focus has been on prediction
of the spatial or spatio-temporal variation in a response surface,
rather than on
estimation of model parameters. In problems of this kind, where
parameters are not of direct
interest but rather are a means to an end, Bayesian prediction in
conjunction with diffuse
priors is an attractive strategy, as its predictions naturally accommodate
the effect of parameter uncertainty. Model-based predictions are essentially
nonparametric
smoothers, but embedded within a probabilistic framework. This
encourages the user to
present results in a way that emphasises, rather than hides, their
inherent imprecision.

In
many public health settings, identifying where and when a particular
phenomenon, such as
disease incidence, is likely to have exceeded an agreed intervention
threshold is more
useful than quoting either a point estimate and its standard error or the
statistical significance of departure from a benchmark.

The log-linear formulation is convenient because of the tractable
moment properties of
the log-Gaus\-sian distribution. It also gives the model a natural interpretation
as a multiplicative
decomposition of the overall intensity into deterministic and
stochastic components. However, it can lead to very highly skewed
marginal distributions,
with large patches of near-zero intensity interspersed with sharp\vadjust{\goodbreak}
peaks. Within the
Monte Carlo inferential framework, there is no reason why other, less
severe transformations
from $\IR$ to $\IR^+$ should not be used.

Two areas of current methodological research are the formulation of
models and methods
for principled analysis of multiple data streams that include data of variable
quality from nontraditional
sources, and the further development of robust computational algorithms
that can
deliver reliable inferences for
problems of ever-increasing complexity.

Our general approach reflects a continuing trend in applied statistics
since the 1980s.
The explosion in the development of computationally intensive methods
and associated
complex stochastic models has encouraged a move away from a methods-based
classification of the statistics discipline and towards
a multidisciplinary,
problem-based focus in which statistical method (singular) is
thoroughly embedded within
scientific method.

\section*{Acknowledgements}

We thank the Department of Environmental and Cancer Epidemiology in the
National Center For Epidemiology (Spain) for providing aggregated data
from the
Castile-La Mancha region for permission to use the Spanish lung cancer data.

The leptospirosis study described in Section~\ref{sec:synthesis} is funded by a USA
National Science Foundation grant, with Principal Investigator
Professor Albert Ko
(Yale University School of Public Health).
This work was supported by the UK Medical Research Council\break  (Grant
number G0902153).

\begin{supplement}[id=suppA]
\stitle{Supplementary materials for ``Spatial and spatio-temporal log-Gaussian Cox
processes: Extending the geostatistical paradigm''\\}
\slink[doi]{10.1214/13-STS441SUPP} 
\sdatatype{.pdf}
\sfilename{sts441\_supp.pdf}
\sdescription{This material contains mixing, convergence and inferential diagnostics
for all of the examples in the main article and is also available from
\href{http://www.lancs.ac.uk/staff/taylorb1/statsciappendix.pdf}{http://www.lancs.ac.uk/}
\href{http://www.lancs.ac.uk/staff/taylorb1/statsciappendix.pdf}{staff/taylorb1/statsciappendix.pdf}.}
\end{supplement}


%

\end{document}